\documentclass[english]{article}
\usepackage[T1]{fontenc}
\usepackage[latin9]{inputenc}
\usepackage{geometry}
\usepackage{units}
\usepackage{amsmath}
\usepackage{amssymb}
\usepackage{wasysym}
\usepackage{babel}
\begin{document}
\title{Series Solution for Interaction of Scalar Plane Wave with \\
Spatially Decaying Gravitational Wave }
\author{Jesse Elder and Todd Fugleberg\thanks{Electronic address: fuglebergt@brandonu.ca}\\
Department of Physics and Astronomy, Brandon University\\
John R. Brodie Science Centre, 270 - 18th Street \\
Brandon, Manitoba R7A 6A9}
\maketitle
\begin{abstract}
In this paper we present the power series solution of the Klein-Gordon
equation in the spacetime background of a gravitational wave with
amplitude that decays with distance from the source. The resulting
solution describes the interaction of a scalar plane wave travelling
in an arbitrary direction relative to the direction of propagation
of the gravitational wave. This solution has the unexpected property
that as the scalar wave approaches collinearity with the gravitational
wave there is a rapid transition in the form of the solution. The
solution in the collinear limit exhibits a resonance phenomenon which
distinguishes these results from previous analyses involving plane
gravitational wave backgrounds. We discuss in detail the similarities
and differences between the solutions for plane gravitational waves
and gravitational waves with amplitude that decreases with distance
from the source. We give an argument that this solution of the Klein-Gordon
equation only describes the interaction of a gravitational wave with
a scalar wave and that the gravitational wave will not produce a scalar
waveform in a vacuum. The interaction between the gravitational and
scalar waves lead to both sinusoidal time-dependent fluctuations in,
and time-independent enhancement of, the scalar current in the direction
of the gravitational wave. Finally, we discuss the possibility of
observable effects of this interaction.
\end{abstract}

\section{Introduction}

Gravitational waves, first predicted by Einstein the year after the
publication of his general theory of relativity, have long been the
subject of great interest to physicists. This interest has intensified
and widened to the general public since the first direct detection
of gravitational waves in 2015 \cite{Abbot et al} . The lag between
indirect confirmation of the existence of gravitational waves in 1982
\cite{Taylor Weisberg} and direct detection was largely due to the
extremely small amplitude of gravitational waves. The amplitude of
a gravitational wave is determined by the gravitational wave strain
which is a dimensionless quantity that describes how much spatial
distances are stretched and compressed by the gravitational wave.
The gravitational strain detectable by the current generation of gravitational
wave detectors is roughly in the range of $10^{-19}\,-\,10^{-23}$
depending on the frequency \cite{Third Run Second Half}. The extremely
small size of the observed strains is partly due to the general relativistic
analysis of their production and partly due to the very large distances
to the gravitational wave sources. As a result there has been much
interest in other observable effects of gravitational waves, both
before and after the direct detection. The current generation of earth-based
gravitational wave detectors have limits not only on the amplitude
but also on the frequency of gravitational waves that they can detect.
Projected space-based detectors will enlarge the parameter space for
detection of gravitational waves \cite{LISA}, but are many years
in the future. Therefore any analysis that leads to observable effects
of gravitational waves will be of value in expanding our knowledge
of this phenomenon.

One interesting avenue is the study of the interaction of gravitational
waves with other physical fields. There have been many recent publications
\cite{Jones - Scalar Production}-\cite{Brandenberger} considering
the solution of the Klein-Gordon, Dirac and Maxwell's equations in
the spacetime background of a plane gravitational wave. These publications
discuss interesting predictions for possible observable effects of
the interaction of gravitational waves with different types of physical
fields. The subject of this paper is the solution of the massless
Klein-Gordon equation, using a power series in terms of inverse distance
from the source, in the spacetime background of a gravitational wave
with amplitude that decreases like $\nicefrac{1}{r}$ with distance
from the source. This solution will describe the effect of the gravitational
wave on a plane wave of a massless scalar field travelling in an arbitrary
direction relative to the gravitational wave.

At first glance, it would seem unnecessary to include any distance
dependence on the amplitude of gravitational waves due to the large
distances from real astrophysical sources and the much smaller size
of any practically realizable detector. Gravitational waves at earth
from sources at distances in the range of currently observed sources
$\left(40\,Mpc-445\,Gpc\right)$ \cite{Third Run Second Half,Third Run First Half,First and Second Runs}
will be planar to a very good approximation. We will show in this
paper, however, that the distance dependence of the amplitude of the
gravitational wave will have significant implications in the limit
where the scalar wave is collinear with (travelling in the same direction
as) the gravitational wave. For scalar waves which are not close to
collinearity, however, the results of this analysis agree very well
with previous results.

We only considered a massless scalar field in this analysis to limit
the length of this paper. Even though there are no massless scalar
fields that we know of, at least in the current epoch of the universe's
evolution, the knowledge gained in this analysis will be useful in
applying the same procedure to fermionic and EM fields. It should
be noted that, this extension to other physical fields has been done
(\cite{Jones -  LF EM}-\cite{Dasgupta - EM}, \cite{Dasgupta - Yang Mills-1})
for plane gravitational waves. We also restricted this analysis to
the $+$-polarization state of quadrupolar gravitational waves. The
results derived herein can easily be extended to the $\times$-polarization
state and generalized to a linear combination of the two. Finally,
we did not consider the back-reaction of this interaction on the gravitational
field itself.

In the next section of the paper, we present the general solution
of the massless Klein-Gordon (KG) equation in a flat space background
in light-cone coordinates and give necessary relations for components
of the wave-vector in light-cone coordinates. In the Section \ref{sec:Gravitational-Wave-Background}
we derive the form of the massless KG equation in the spacetime background
of a  gravitational wave with amplitude that decays with distance
from the source. In Section \ref{sec:Collinear-Case} we discuss the
solution of this KG equation in the collinear limit - where the scalar
wave is moving in the same direction as the gravitational wave which
was taken to be along the $z-$axis. In Section \ref{sec:Special-Non-Collinear-Case}
we discuss the solution of the KG equation in a special noncollinear
case where the propagation direction of the scalar wave has a component
perpendicular to the $z-$axis which is midway between the $x$ and
$y$ axes\footnote{For $+$-polarized gravitational waves the $x$ and $y$ axes are
the directions of maximum stretch/compression of space}. In Section \ref{sec:General-Non-Collinear-Case} we present the
results of the solution in the general case where the scalar wave
is travelling in an arbitrary direction relative to the propagation
direction of the gravitational wave. Some of the details of the analysis
are given in Appendix A. In Section \ref{sec:Four-Current} we discuss
the four-current derived from the general solution with some of the
details given in Appendix B. In Section \ref{sec:No-Particle-Production}
we argue that the solutions discussed in this paper only involve the
interaction of the gravitational wave with a scalar plane wave and
that a gravitational wave will not produce any scalar excitations
in the absence of this plane wave. In Section \ref{sec:Physical-Implications}
we discuss a few different physical implications and the possibility
of their observation. Finally in Section \ref{sec:Conclusion} we
discuss the conclusions that we draw from this analysis and possibilities
for future research.

\section{Klein-Gordon Equation on Light Cone\label{sec:Klein-Gordon-Equation}}

In Cartesian coordinates, the massless KG equation in flat space:
\begin{equation}
\square\Psi=\left(-\partial_{t}^{2}+\partial_{x}^{2}+\partial_{y}^{2}+\partial_{z}^{2}\right)\Psi=0\label{eq:flat KG eqn}
\end{equation}
has the plane wave solution:
\begin{equation}
\Psi\left(t,x,y,z\right)={\cal A}e^{i\vec{K}\cdot\vec{R}}={\cal A}e^{-i\omega t}e^{ik_{x}x}e^{ik_{y}y}e^{ik_{z}z},\label{eq:quantum plane wave}
\end{equation}
with null four-momentum satisfying:
\begin{equation}
-\omega^{2}+k_{x}^{2}+k_{y}^{2}+k_{z}^{2}=0.\label{eq:lightcone condition cartesian}
\end{equation}
Using light cone coordinates defined by:
\begin{eqnarray}
u=t-z & \mbox{\qquad} & z=\frac{v-u}{2},\label{eq:light cone coordinates}\\
v=t+z & \mbox{\qquad} & t=\frac{v+u}{2},\nonumber 
\end{eqnarray}
we can write:
\begin{equation}
\vec{K}\cdot\vec{R}=-\frac{\left(k_{z}+\omega\right)}{2}\,u+\frac{\left(k_{z}-\omega\right)}{2}\,v+k_{x}x+k_{y}y,\label{eq:k dot R}
\end{equation}
and make the following definitions:
\begin{eqnarray}
k_{u}=-\frac{k_{z}+\omega}{2} & \qquad & k_{v}+k_{u}=-\omega\label{eq:light cone momenta coords definitions}\\
k_{v}=\frac{k_{z}-\omega}{2} & \qquad & k_{v}-k_{u}=k_{z}.\nonumber 
\end{eqnarray}

In light cone coordinates the KG equation becomes:
\begin{equation}
\left(-4\partial_{u}\partial_{v}+\partial_{x}^{2}+\partial_{y}^{2}\right)\Psi=0\label{eq:KG in light cone}
\end{equation}
with plane wave solution:

\begin{equation}
\Psi\left(u,x,y,v\right)={\cal A}e^{i\vec{K}\cdot\vec{R}}={\cal A}e^{ik_{u}u}e^{ik_{v}v}e^{ik_{x}x}e^{ik_{y}y}\label{eq:plane wave in light cone}
\end{equation}
where the null condition becomes:
\begin{equation}
-4k_{u}k_{v}+k_{x}^{2}+k_{y}^{2}=0.\label{eq:light cone condition uv coords}
\end{equation}

In the analyis of this paper we take the direction of propagation
of the gravitational wave to be the $z-$direction and the direction
of the maximum stretch/compression of spatial distances for the $+$-polarized
wave to be the $x$ and $y$ directions. The spatial direction of
propagation of the scalar plane wave can be specified with respect
to this coordinate system by:
\begin{eqnarray}
k_{z}=\omega\cos\theta & \qquad k_{x}=\omega\sin\theta\cos\phi\qquad & k_{y}=\omega\sin\theta\sin\phi\label{eq:Cartesian momentum components}
\end{eqnarray}
for $\theta\in\left[0,\pi\right]$ and $\phi\in\left[0,2\pi\right]$.
In terms of these variables, the lightcone momentum components can
be written:
\begin{eqnarray}
k_{u}=-\omega\frac{\left(1+\cos\theta\right)}{2} &  & k_{v}=-\omega\frac{\left(1-\cos\theta\right)}{2}.\label{eq:lightcone momentum components}
\end{eqnarray}

In later sections the following limiting cases will be relevant:
\begin{eqnarray}
\lim_{\theta\rightarrow0}\frac{k_{x}^{2}+k_{y}^{2}}{4\,k_{v}} & = & \lim_{\theta\rightarrow0}\frac{\omega^{2}\sin^{2}\theta}{-2\omega\,\left(1-\cos\theta\right)}=-\omega,\label{eq:limiting A}
\end{eqnarray}
\begin{eqnarray}
\lim_{\theta\rightarrow0}\frac{k_{y}^{2}-k_{x}^{2}}{2\,k\,k_{v}} & = & \lim_{\theta\rightarrow0}\frac{\omega^{2}\sin^{2}\theta\left(\sin^{2}\phi-\cos^{2}\phi\right)}{-\omega\,k\,\left(1-\cos\theta\right)}=-2\cos\left(2\phi\right)\frac{\omega}{k}.\label{eq:limiting E coord. sing.}
\end{eqnarray}
These limits are both finite and nonzero.

\section{Gravitational Wave Background\label{sec:Gravitational-Wave-Background}}

The Klein-Gordon (KG) equation in a curved spacetime is given by \cite{Birrell Davies}:

\begin{equation}
\frac{1}{\sqrt{-\left|g_{a\beta}\right|}}\partial_{\mu}\left(g^{\mu\nu}\sqrt{-\left|g_{a\beta}\right|}\,\partial_{\nu}\phi\right)=0.\label{eq:Full KG}
\end{equation}
In a Cartesian coordinate system centred at the source of the gravitational
wave, the metric for a gravitational wave with $+$-polarization and
travelling in the $z-$direction can be written:
\begin{equation}
ds^{2}=-dt^{2}+dz^{2}+f\left(u,v\right)^{2}dx^{2}+g\left(u,v\right)^{2}dy^{2},\label{eq:grav wave spacetime interval}
\end{equation}
where the functions $f\left(u,v\right)$ and $g\left(u,v\right)$
determine the exact form of the gravitational wave. It should be noted
that in the transverse-traceless gauge this assumption is valid at
any point on the surface of a sphere around the source \cite{Schutz}.

In light-cone coordinates, $\left(u,x,y,v\right)$, it can easily
be shown that the metric corresponding to (\ref{eq:grav wave spacetime interval})
is given by:
\begin{equation}
g_{\alpha\beta}=\left(\begin{array}{cccc}
0 & 0 & 0 & -\frac{1}{2}\\
0 & f^{2}\left(u,v\right) & 0 & 0\\
0 & 0 & g^{2}\left(u,v\right) & 0\\
-\frac{1}{2} & 0 & 0 & 0
\end{array}\right),\label{eq:metric}
\end{equation}
and the exact form of the inverse metric is:
\begin{equation}
g^{\alpha\beta}=\left(\begin{array}{cccc}
0 & 0 & 0 & -2\\
0 & \nicefrac{1}{f^{2}\left(u,v\right)} & 0 & 0\\
0 & 0 & \nicefrac{1}{g^{2}\left(u,v\right)} & 0\\
-2 & 0 & 0 & 0
\end{array}\right).\label{eq:inv metric}
\end{equation}
It is well known \cite{Schutz} that solutions of the linearized Einstein
equation have the form:
\begin{eqnarray}
f\left(u,v\right)=1+\frac{2he^{iku}}{\left(v-u\right)}=1+\frac{h}{z}e^{iku} &  & g\left(u,v\right)=1-\frac{2he^{iku}}{\left(v-u\right)}=1-\frac{h}{z}e^{iku}.\label{eq:f and g functions}
\end{eqnarray}
The quantity $\nicefrac{h}{z}$ is the dimensionless gravitational
strain at distance $z=\frac{v-u}{2}$ from the source of the gravitational
wave. It should be noted at this point that the non-spherical nature
of the gravitational wave is entirely contained in the variable $h$.
The value of $h$ will depend on the angular position of the observation
point relative to the axis of symmetry of the source but will be essentially
constant over the transverse extent of the wave at even relatively
small astronomical distances.

In the gravitational background specified above it can be shown that
the massless KG equation in light cone coordinates is:
\begin{equation}
F(u,v)\,\partial_{u}\partial_{v}\Psi-ikG(u,v)\,\partial_{v}\Psi+\frac{1}{2}J(u,v)\left(\partial_{u}-\partial_{v}\right)\Psi-\frac{1}{4}\,\left(H_{x}(u,v)\partial_{x}^{2}+H_{y}(u,v)\partial_{y}^{2}\right)\Psi=0,\label{eq:our DE}
\end{equation}
where:
\begin{eqnarray}
F(u,v) & = & \left(1-\frac{4h^{2}}{\left(v-u\right)^{2}}\,e^{2iku}\right)^{2},\nonumber \\
G(u,v) & = & \left(1-\frac{4h^{2}}{\left(v-u\right)^{2}}\,e^{2iku}\right)\frac{4h^{2}}{\left(v-u\right)^{2}}\,e^{2iku},\nonumber \\
J(u,v) & = & \left(1-\frac{4h^{2}}{\left(v-u\right)^{2}}\,e^{2iku}\right)\frac{8h^{2}}{\left(v-u\right)^{3}}\,e^{2iku},\label{eq:coefficient functions}\\
H_{x}(u,v) & = & \left(1-\frac{2h}{\left(v-u\right)}\,e^{iku}\right)^{2},\nonumber \\
H_{y}(u,v) & = & \left(1+\frac{2h}{\left(v-u\right)}\,e^{iku}\right)^{2}.\nonumber 
\end{eqnarray}
This equation can be compared with the KG equation in \cite{Jones - Scalar Production}
for the propagation of a scalar wave in the background of a plane
gravitational wave. The first two terms in (\ref{eq:our DE}) are
equivalent to the respective terms in Eq. (8) in \cite{Jones - Scalar Production}
(see also Eq. (\ref{eq:Preston KG}) in next section) with the substitution
$\nicefrac{h}{z}=\nicefrac{2h}{(v-u)}\rightarrow h_{+}$. We will
show in Section \ref{sec:Special-Non-Collinear-Case} that the last
term in (\ref{eq:our DE}) is equivalent to the last term in Eq. (8)
in \cite{Jones - Scalar Production} in the special case that they
consider. The term involving $J\left(u,v\right)$ in Eq. (\ref{eq:our DE})
arises as a result of the $z-$dependence of the gravitational wave
amplitude and has no corresponding term in \cite{Jones - Scalar Production}.
The expressions for $F(u,v),\,G(u,v),\,H_{x}(u,v)$ and $H_{y}(u,v)$
involve powers of $\nicefrac{2he^{iku}}{\left(v-u\right)}$ which
come from $f\left(u,v\right),\,g\left(u,v\right)\,\partial_{u}f\left(u,v\right)$
and $\partial_{u}g\left(u,v\right)$ terms in Eq. (\ref{eq:Full KG}).
The expression for $J\left(u,v\right)$ includes an overall factor
of $\nicefrac{8h^{2}}{\left(v-u\right)^{3}}\,e^{2iku}$ which comes
from $\partial_{u}f\left(u,v\right),\,\partial_{v}f\left(u,v\right),\partial_{u}g\left(u,v\right)$
and $\partial_{v}g\left(u,v\right)$ terms because of the $z-$dependence
of the amplitude of the gravitational wave given by $\nicefrac{2he^{iku}}{\left(v-u\right)}=\nicefrac{he^{iku}}{z}$.
Therefore the KG equation in Eq. (\ref{eq:our DE}) generalizes from
that in \cite{Jones - Scalar Production} due to the inclusion of
the term involving $J\left(u,v\right)$ and the $\unit{1}/\left(v-u\right)$
dependence in the other terms.

It is not essential to do the analysis in terms of lightcone coordinates,
but it simplifies some of the discussion and facilitates comparison
to previous results. As well, the complexity of the analysis would
be similar in another coordinate system and the results would be essentially
the same.

\section{Collinear Case\label{sec:Collinear-Case}}

We will now discuss the analysis of the collinear case $\left(\theta=0\right)$
before discussing more general cases for two reasons. First, we will
show that the collinear case is qualitatively different from the noncollinear
case and presenting this analysis separately will show that the collinear
results are not artifacts of the general analysis. Secondly, analyzing
the collinear case seperately allows a clearer illustration of where
the differences between the approximate noncollinear solution and
the collinear solution come from.

Assume that in flat spacetime there exists a plane wave of a massless
scalar field propagating in the $z-$direction (\emph{ie. $\theta=0$
}and \emph{$k_{x}=k_{y}=k_{v}=0$}). We can determine how the gravitational
wave travelling in the $z-$direction will affect this scalar wave
by solving the Klein-Gordon equation (\ref{eq:our DE}) in this spacetime
background. In the collinear case we can assume that the wave function
is independent of $x$ and $y$, but will depend on $u$ and $v$
due to the $F(u,v),\,G(u,v)$ and $J(u,v)$ factors in the KG equation.
In this case the KG equation becomes:
\begin{equation}
F(u,v)\,\partial_{u}\partial_{v}\Psi\left(u,v\right)-ikG(u,v)\,\partial_{v}\Psi\left(u,v\right)+\frac{1}{2}J(u,v)\left(\partial_{u}-\partial_{v}\right)\Psi\left(u,v\right)=0.\label{eq:KG collinear}
\end{equation}
This equation cannot be solved analytically in an obvious way, so
we sought a power series solution in $\nicefrac{1}{z}=\nicefrac{2}{\left(v-u\right)}$
to this differential equation.

With foresight, we assumed a solution of the form: 
\begin{equation}
\Psi\left(u,v\right)={\cal A}e^{iAu}\left(1+\frac{4B\,h^{2}e^{2iku}}{\left(v-u\right)^{2}}+\frac{8i\,C\,h^{2}e^{2iku}}{\left(v-u\right)^{3}}+\frac{16\,D\,h^{4}e^{4iku}}{\left(v-u\right)^{4}}\right),\label{eq:collinear assumed solution}
\end{equation}
substituted this into Eq. (\ref{eq:KG collinear}) and kept all nonzero
terms up to order $1/z^{4}\propto\nicefrac{1}{\left(v-u\right)^{4}}$
to get the following equation:
\begin{equation}
\frac{8ih^{2}e^{2iku}\left(A-2B\left(A+2k\right)\right)}{\left(v-u\right)^{3}}-\frac{48h^{2}e^{2iku}\left(B-C\left(A+2k\right)\right)}{\left(v-u\right)^{4}}=0.\label{eq:DE terms collinear case}
\end{equation}
This equation can be solved by solving the two constraint equations:
\begin{eqnarray}
A-2B\left(A+2k\right)=0 & \mbox{ \hspace{0.5in}\& \hspace{0.5in}} & B-C\left(A+2k\right)=0,\label{eq:collinear conditions}
\end{eqnarray}
for the coefficients $B$ and $C$ with the results:
\begin{eqnarray}
B=\frac{A}{2\left(A+2k\right)} & \mbox{ \hspace{0.5in}\& \hspace{0.5in}} & C=\frac{A}{2\left(A+2k\right)^{2}},\label{eq:collinear coeff solutions}
\end{eqnarray}
with no constraint on the value of $D$. We have included the $D$
term in this case for consistency with later cases. Finally, in order
for $\Psi\left(u,v\right)$ to match the plane wave solution in the
limit of no gravitational wave ($h\rightarrow0$), we know from Eq.
(\ref{eq:lightcone momentum components}) that $A=-\omega$ in the
collinear limit, $\left(\theta\rightarrow0\right)$. Therefore the
power series solution of (\ref{eq:KG collinear}) up to order $1/z^{4}$
is:
\begin{equation}
\Psi\left(u,v\right)={\cal A}e^{-i\omega u}\left(1-\frac{2\omega\,h^{2}e^{2iku}}{\left(-\omega+2k\right)\left(v-u\right)^{2}}-\frac{4i\,\omega\,h^{2}e^{2iku}}{\left(-\omega+2k\right)^{2}\left(v-u\right)^{3}}+\frac{16\,D\,h^{4}e^{4iku}}{\left(v-u\right)^{4}}\right).\label{eq:collinear solution}
\end{equation}

It is useful, at this point to check the consistency of this result
with previous results. In \cite{Dasgupta - Scalar Fermion} they considered
perturbations of the wave function, up to first order in the gravitational
strain, caused by a plane gravitational wave. They found that these
perturbations vanish in the collinear limit. In our analysis, the
gravitational strain is distance dependent, $\nicefrac{h}{z}\propto\nicefrac{h}{\left(v-u\right)}$,
and the strain dependent contributions to Eq. (\ref{eq:collinear solution})
occur at second and higher order. Therefore, the collinear results
for our gravitational wave model are consistent with the results in
\cite{Dasgupta - Scalar Fermion} for a plane gravitational wave.

Comparing with the results of \cite{Jones - Scalar Production} requires
a little more discussion. For a plane gravitational wave the authors
of \cite{Jones - Scalar Production} considered the Klein-Gordon equation:
\begin{equation}
\left(4F(u)\,\partial_{u}\partial_{v}-4ikG(u)\,\partial_{v}+\,H(u)\left(\partial_{x}^{2}+\partial_{y}^{2}\right)\right)\phi=0,\label{eq:Preston KG}
\end{equation}
where $F(u)$ and $G(u)$ are equivalent to $F(u,v)$ and $G(u,v)$
in (\ref{eq:coefficient functions}) with the substitution $\nicefrac{h}{z}=\nicefrac{2h}{\left(v-u\right)}\rightarrow h_{+}$
and, as we will show in the next section, $H(u)$ is equivalent to
$H_{x}(u,v)+H_{y}(u,v)$ under the same substitution. The difference
in signs of the terms involving $\partial_{x}^{2}$ and $\partial_{y}^{2}$
between Eq'ns (\ref{eq:our DE}) \& (\ref{eq:Preston KG}) is due
to the differing coordinate definitions where the authors of \cite{Jones - Scalar Production}
used $u=z-t$ and we used $u=t-z$ as in \cite{Schutz}. The authors
of \cite{Jones - Scalar Production} assumed a separable form of the
solution: 
\begin{equation}
\phi=U(u)e^{ik_{v}v}e^{ik_{x}x}e^{ik_{y}y}.\label{eq:Preston separable solution}
\end{equation}
In the collinear limit $\left(k_{x},\,k_{y},\,k_{v}\rightarrow0\right)$,
this leads to $\phi\equiv U(u)$, and their KG equation vanishes identically.
Any function of $u=t-z$ will satisfy the collinear limit of their
KG equation\footnote{This is analogous to the well known fact that any twice differentiable
function of $(x-vt)$ will satisfy the 1D wave equation.}. Therefore, for a plane gravitational wave, the collinear form of
their KG equation does not constrain the solution in the collinear
limit. On the other hand, in \cite{Jones - Scalar Production} they
solve the general form of their KG equation and their general solution
has a specific collinear limiting form. In our case, involving a  gravitational
wave with $z-$dependent amplitude, the solution of the collinear
KG equation in Eq. (\ref{eq:KG collinear}) has a particular form
which the general solution must match in the collinear limit. We will
leave discussion of the physical implications of the power series
solution in Eq. (\ref{eq:collinear solution}) until Section \ref{sec:Physical-Implications}.

Before moving on, however, we need to comment on the order of the
power series approximation. The functions $F(u,v)$ and $G(u,v)$
terms in the KG equation in (\ref{eq:KG collinear}) are defined in
(\ref{eq:coefficient functions}) and have highest order terms proportional
to $\nicefrac{1}{\left(v-u\right)^{4}}$ and are equivalent to the
$F(u)$ and $G(u)$ terms from \cite{Jones - Scalar Production}.
The terms involving $H_{x}(u,v)$ and $H_{y}(u,v)$ also have highest
order $\nicefrac{1}{\left(v-u\right)^{4}}$ and, as we shall show
in the next section, the combination of these terms is equivalent
to the term in Eq. (\ref{eq:Preston KG}) involving $H(u)$. Therefore
to facilitate comparison with the results of \cite{Jones - Scalar Production}
it is necessary to keep terms in the expansion up to at least $\nicefrac{1}{\left(v-u\right)^{4}}$. 

The reader might question why we include terms up to fourth order
in $\nicefrac{1}{\left(v-u\right)}$ in our series solution, when
according to Eq. (\ref{eq:f and g functions}), the gravitational
wave solution seems to be first order in $\nicefrac{1}{\left(v-u\right)}$.
It should be clear, however, that using these expressions for $f\left(u,v\right)$
and $g\left(u,v\right)$ in Eq'n (\ref{eq:metric}) leads to a metric
which is second order in $\nicefrac{1}{\left(v-u\right)}$ and then
using this $g_{\mu\nu}$ and the exact inverse $g^{\mu\nu}$ in Eq.
(\ref{eq:Full KG}) naturally leads to an equation which is fourth
order in $\nicefrac{1}{\left(v-u\right)}$. Using a nonlinear gravitational
wave solution instead of Eq. (\ref{eq:f and g functions}) would be
preferable but would extend an already long analysis even more. For
this reason, as is commonly done, we choose to work in the spacetime
background of a linearized gravitational wave.

\section{Special Noncollinear Case\label{sec:Special-Non-Collinear-Case}}

We will see in the next section that the solution for the KG equation
(\ref{eq:our DE}) for a plane wave of a massless scalar field travelling
in an arbitrary direction is very complex. In order to clearly illustrate
some of the details of this analysis we will first discuss the solutions
of a special noncollinear case in detail. The special case we consider
has $\phi=\frac{\pi}{4},\frac{5\pi}{4}$ $\left(ie.\:k_{x}=k_{y}\right)$
and is the same special case that was considered in \cite{Jones - Scalar Production}
for a plane gravitational wave which allowed them to exactly solve
their KG equation. This analysis will determine the effect of the
spatially dependent gravitational wave on a scalar plane wave which
is not collinear with the gravitational wave and has perpendicular
components midway between the $x$ and $y$ directions. In this case,
since $k_{x}=k_{y}$, the derivatives of the wavefunction with respect
to $x$ and $y$ are the same ($\partial_{x}^{2}\Psi=\partial_{y}^{2}\Psi$)
which means the $x$ and $y$ dependent term of the KG equation (\ref{eq:our DE})
becomes:
\begin{eqnarray}
\frac{1}{4}\,\left(H_{x}(u,v)\partial_{x}^{2}+H_{y}(u,v)\partial_{y}^{2}\right)\Psi & = & \frac{1}{4}\,\left(\left(1-\frac{2h}{\left(v-u\right)}\,e^{iku}\right)^{2}+\left(1+\frac{2h}{\left(v-u\right)}\,e^{iku}\right)^{2}\right)\partial_{x}^{2}\Psi\nonumber \\
 & = & \frac{1}{2}\,\left(1+\frac{4h^{2}}{\left(v-u\right)^{2}}\,e^{2iku}\right)\partial_{x}^{2}\Psi\nonumber \\
 & \equiv & \frac{1}{2}\,H(u,v)\left(\frac{\partial_{x}^{2}\Psi+\partial_{y}^{2}\Psi}{2}\right).\label{eq:definition of H}
\end{eqnarray}
In this special case the Klein-Gordon equation becomes:
\begin{equation}
F(u,v)\,\partial_{u}\partial_{v}\Psi-ikG(u,v)\,\partial_{v}\Psi+\frac{1}{2}J(u,v)\left(\partial_{u}-\partial_{v}\right)\Psi-\frac{1}{4}H(u,v)\left(\partial_{x}^{2}\Psi+\partial_{y}^{2}\Psi\right)=0\label{eq:special noncollinear KG}
\end{equation}
which is now seen to be analogous to Eq. (\ref{eq:Preston KG}) except
for the $J(u,v)$ term, the $z-$dependence in $F(u,v),\,G(u,v)$
and $H(u,v)$, and the sign of the $H(u,v)$ term due to different
definitions of the coordinate $u$. With foresight, we assume a solution
of the form:

\begin{eqnarray}
\Psi\left(u,x,y,v\right) & = & {\cal A}e^{iAu}e^{ik_{v}v}e^{ik_{x}x}e^{ik_{y}y}\left[1+\frac{4B_{1}\,h^{2}e^{2iku}}{\left(v-u\right)^{2}}+\frac{8i\,C\,h^{2}e^{2iku}}{\left(v-u\right)^{3}}+\frac{16\,D\,h^{4}e^{4iku}}{\left(v-u\right)^{4}}\right.\nonumber \\
 &  & \hspace{1.2in}\left.+\frac{4\,B_{2}\,e^{2iku}}{\left(v-u\right)^{2}}\left(1+\frac{2iB_{3}}{k_{v}\left(v-u\right)}+\frac{4\left(B_{4}k_{v}+B_{5}\right)}{k_{v}^{2}\left(v-u\right)^{2}}\right)^{-1}\right]\negthickspace.\label{eq:special noncollinear solution}
\end{eqnarray}
This solution is not a pure power series, but if one were to (for
$k_{v}\neq0$) do a power series expansion of the inverse expression
in the last term, the whole solution would be a power series of order
$\nicefrac{1}{(v-u)^{4}}$. Doing so would be a good approximation
for most values of $k_{v}$, but would not be valid near the collinear
limit when $\frac{1}{k_{v}\left(v-u\right)}\gtrsim1$ and would lead
to a solution which diverges in the collinear limit $\left(k_{v}\rightarrow0\right)$
. In what follows, we will discuss how this form of the assumed solution
avoids this divergence. A similiar issue will arise later when we
discuss the four-current in Section \ref{sec:Four-Current} and we
will discuss this issue further at that point. It can easily be seen
that the collinear limit ($k_{v},\,k_{x},\,k_{y}\rightarrow0$) of
the assumed form in (\ref{eq:special noncollinear solution}) is the
same as Eq. (\ref{eq:collinear assumed solution}). 

In order to find the values of the unspecified coefficients in (\ref{eq:special noncollinear solution})
the following procedure was used: 1) substitute assumed solution (\ref{eq:special noncollinear solution})
into Eq. (\ref{eq:special noncollinear KG}), 2) multiply all resulting
terms by $\left(1+\frac{2iB_{3}}{k_{v}\left(v-u\right)}+\frac{4\left(B_{4}k_{v}+B_{5}\right)}{k_{v}^{2}\left(v-u\right)^{2}}\right)^{2}$
to cancel factors in denominators of derivatives of the last term
in Eq. (\ref{eq:special noncollinear solution})\footnote{One might worry that this could lead to a problem in the collinear
limit, because this factor diverges when $k_{v}\rightarrow0$. The
validity of this step will rely on fact that this analysis leads to
the correct solution in the collinear limit.}, and 3) combine terms with same powers of $\nicefrac{1}{\left(v-u\right)}$
and $e^{iku}$ keeping terms up to order $\nicefrac{1}{(v-u)^{4}}$.
The terms arising from these three steps are as follows:

\begin{eqnarray}
\mbox{indep. of }h\,e^{iku}: &  & \left(-2A\,k_{v}+\frac{k_{x}^{2}}{2}+\frac{k_{y}^{2}}{2}\right)\left(1+\frac{2iB_{3}}{k_{v}\left(v-u\right)}+\frac{4\left(B_{4}k_{v}+B_{5}\right)}{k_{v}^{2}\left(v-u\right)^{2}}\right)^{2}\nonumber \\
 &  & \equiv\alpha\left(1+\frac{2iB_{3}}{k_{v}\left(v-u\right)}+\frac{4\left(B_{4}k_{v}+B_{5}\right)}{k_{v}^{2}\left(v-u\right)^{2}}\right)^{2},\label{eq:constant term in special case}\\
\frac{1}{(v-u)^{2}}\:\mbox{terms}: &  & \frac{4h^{2}e^{2iku}}{(v-u)^{2}}\:k_{v}\left[2\left(3A+k\right)-4\left(B_{1}+B_{2}\right)k+\left(1+B_{1}+B_{2}\right)\frac{\alpha}{k_{v}}\right]\nonumber \\
 &  & \equiv\frac{4h^{2}e^{2iku}}{(v-u)^{2}}\:\delta,\label{eq:second order term in special case}\\
\frac{1}{(v-u)^{3}}\:\mbox{terms}: &  & -\frac{8ih^{2}e^{2iku}}{(v-u)^{3}}\,\left[2\left(B_{1}+B_{2}\right)\left(A+2k-k_{v}\right)-A+k_{v}-4k\left(B_{2}B_{3}-Ck_{v}\right)\begin{array}{c}
\,\\
\,
\end{array}\right.\hspace{-2em}\nonumber \\
 &  & \left.\mbox{\hspace{1in}}-2B_{3}\frac{\delta}{k_{v}}+\left(B_{2}B_{3}-Ck_{v}\right)\frac{\alpha}{k_{v}}\right]\nonumber \\
 &  & \equiv-\frac{8ih^{2}e^{2iku}}{(v-u)^{3}}\:\epsilon,\label{eq:third order term in special case}\\
\frac{e^{2iku}}{(v-u)^{4}}\:\mbox{terms}: &  & -\frac{16h^{2}e^{2iku}}{(v-u)^{4}}\left[3\left(A+2k-k_{v}\right)\left(\frac{B_{2}B_{3}}{k_{v}}-C\right)-4B_{2}k\left(\frac{B_{3}^{2}}{k_{v}}+B_{4}+\frac{B_{5}}{k_{v}}\right)\begin{array}{c}
\,\\
\,
\end{array}\right.\nonumber \\
 &  & \hspace{1in}+3\left(B_{1}+B_{2}\right)+B_{2}\left(\frac{B_{3}^{2}}{k_{v}}+B_{4}+\frac{B_{5}}{k_{v}}\right)\frac{\alpha}{k_{v}}-2B_{3}\frac{\epsilon}{k_{v}}\nonumber \\
 &  & \left.\hspace{1in}-B_{2}\left(3\frac{B_{3}^{2}}{k_{v}}+2B_{4}+2\frac{B_{5}}{k_{v}}\right)\frac{\delta}{k_{v}}\right],\label{eq:first fourth order term in special case}\\
\frac{e^{4iku}}{(v-u)^{4}}\:\mbox{terms}: &  & \frac{16h^{4}e^{4iku}}{(v-u)^{4}}\:k_{v}\left[2\left(B_{1}+B_{2}\right)\left(3A+5k\right)-2\left(A+k+4\,D\,k\right)+\left(B_{1}+B_{2}+D\right)\frac{\alpha}{k_{v}}\right].\label{eq:second fourth order term in special case}
\end{eqnarray}
We have introduced $\alpha,\,\delta,$ and $\epsilon$ to represent
factors from lower order terms in order to write the higher order
terms more succinctly. The first expression is different than the
subsequent expressions as it contains different powers of $\nicefrac{1}{(v-u)}$,
but is kept separate for two reasons: 1) unlike the rest of the expressions,
it does not vanish in the absence of a gravitational wave $\left(h\rightarrow0\right)$,
and 2) the constraint arising from this term is identical to the null
condition for the wavevector in light cone coordinates, Eq. (\ref{eq:light cone condition uv coords}),
and is more fundamental than the other constraints.
\begin{flushleft}
In order to satisfy Eq. (\ref{eq:special noncollinear KG}), each
of these terms must vanish and place constraints on $A$, $B_{1}$
to $B_{5}$, $C$ and $D$. These expressions are not independent
as can be seen by the presence of earlier terms in later terms. The
independent constraint equations derived from these are:
\begin{eqnarray}
-4A\,k_{v}+k_{x}^{2}+k_{y}^{2} & = & 0,\label{eq:special non-collinear 1}\\
k_{v}\left(3A+k-2\left(B_{1}+B_{2}\right)k\right) & = & 0,\label{eq:special non-collinear 2}\\
2\left(B_{1}+B_{2}\right)\left(A+2k-k_{v}\right)-A+k_{v}-4k\left(B_{2}B_{3}-Ck_{v}\right) & = & 0,\label{eq:special non-collinear 3}\\
3\left(A+2k-k_{v}\right)\left(\frac{B_{2}B_{3}}{k_{v}}-C\right)-4kB_{2}\left(\frac{B_{3}^{2}}{k_{v}}+B_{4}+\frac{B_{5}}{k_{v}}\right)+3\left(B_{1}+B_{2}\right) & = & 0,\label{eq:special non-collinear 4}\\
k_{v}\left(\left(B_{1}+B_{2}\right)\left(3A+5k\right)-\left(A+k+4\,D\,k\right)\right) & = & 0.\label{eq:special non-collinear 5}
\end{eqnarray}
There are three other constraints necessary to determine all eight
unknown variables. The first extra constraint is that the collinear
limit of $B_{1}$ agrees with the collinear solution for $B$:
\begin{equation}
\lim_{k_{v}\rightarrow0}B_{1}=\frac{A}{2\left(A+2k\right)}.\label{eq:limiting B1}
\end{equation}
The next two constraints are that $C$ is finite in the collinear
limit and that its limiting value agrees with the collinear solution
for $C$:
\begin{eqnarray}
\lim_{k_{v}\rightarrow0}C<\infty & \hspace{0.5in}\&\hspace{0.5in} & \lim_{k_{v}\rightarrow0}C=\frac{A}{2\,(A+2k)^{2}}.\label{eq:limiting conditions for C}
\end{eqnarray}
The two expressions in (\ref{eq:limiting conditions for C}) might
seem to only constitute a single condition, but as we will show below,
two different coefficients are necessary to satisfy these two conditions.
Therefore we write both conditions in (43) to emphasize that there
are eight constraints on the eight unknown variables. Alternatively,
we could have defined a single variable $B_{4}^{\prime}=B_{4}k_{v}+B_{5}$
and used only the equality in (\ref{eq:limiting conditions for C})
but this approach would have been more difficult.
\par\end{flushleft}

Before solving Eq'ns (\ref{eq:special non-collinear 1}) - (\ref{eq:limiting conditions for C})
we will first consider a digression to illustrate why the final term
in Eq. (\ref{eq:special noncollinear solution}) involving $B_{2}$
to $B_{5}$ is necessary. In the collinear limit ($k_{v},\,k_{x},\,k_{y}\rightarrow0$)
the constraint equations (\ref{eq:special non-collinear 1}), (\ref{eq:special non-collinear 2})
and (\ref{eq:special non-collinear 5}) vanish identically, leaving
only five constraints on eight unknowns. We will show below that $A=-\omega$
in the collinear limit as a result of the light cone condition. Considering
the assumed form in Eq (\ref{eq:special noncollinear solution}) in
the collinear limit ($k_{v}\rightarrow0$), we can see that the term
involving $B_{2}$ to $B_{5}$ vanishes and allows us to set $B_{2}$
to $B_{5}$ to zero. In this case Equations (\ref{eq:special non-collinear 3})
and (\ref{eq:special non-collinear 4}) become:
\begin{eqnarray}
2B_{1}\left(A+2k-k_{v}\right)-A+k_{v}+4Ckk_{v}=0 & \mbox{ \hspace{0.2in}\& \hspace{0.2in}} & 3\left(B_{1}-C\left(A+2k-k_{v}\right)\right)=0\label{eq:collinear limit of noncollinear constraints}
\end{eqnarray}
which in the collinear limit are identical to the collinear constraint
equations in Eq. (\ref{eq:collinear conditions}) with $B_{1}=B$.
Therefore, we have two constraint equations for $B_{1}$ and $C$
which will lead to the same results as in Eq. (\ref{eq:collinear solution})
with $B_{1}=B$ and no constraint on $D$. We now have the necessary
details to make some important observations. 

In the noncollinear case, if we were to eliminate the last term in
Eq. (\ref{eq:special noncollinear solution}) by taking $B_{2}$ to
$B_{5}$ to zero, Equations (\ref{eq:special non-collinear 2}) -
(\ref{eq:special non-collinear 4}) would become:
\begin{eqnarray}
k_{v}\left(3A+k-2B_{1}k\right) & = & 0,\label{eq:new constraints special non-collinear 1}\\
2B_{1}\left(A+2k-k_{v}\right)-A+k_{v}+4Ckk_{v} & = & 0,\label{eq:new constraints special non-collinear 2}\\
-3C\left(A+2k-k_{v}\right)+3B_{1} & = & 0.\label{eq:new constraints special non-collinear 3}
\end{eqnarray}
Recall that $A$ is determined by the constraint equation (\ref{eq:special non-collinear 1}).
When $k_{v}\neq0$, the constraint equations for $B_{1}$ and $C$
are Equations (\ref{eq:new constraints special non-collinear 1})
and (\ref{eq:new constraints special non-collinear 2}) respectively.\footnote{Eq. (\ref{eq:new constraints special non-collinear 3}) is actually
inconsistent with Eq. (\ref{eq:new constraints special non-collinear 2})
but that is just another symptom of the problem we are addressing
here.} When $k_{v}=0$, however, the constraint equations for $B_{1}$ and
$C$ are Equations (\ref{eq:new constraints special non-collinear 2})
and (\ref{eq:new constraints special non-collinear 3}) respectively.
The fact that different equations constrain variables when $k_{v}=0$
than when $k_{v}\neq0$ suggests there is a problem with this analysis
if $B_{2}$ to $B_{5}$ are taken to be zero. Further, solving equations
(\ref{eq:new constraints special non-collinear 1}) and (\ref{eq:new constraints special non-collinear 2})
we obtain the results:
\begin{eqnarray}
B_{1}=\frac{3A+k}{2k} & \hspace{0.5in}\&\hspace{0.5in} & C=\frac{(3A+k)\left(A+2k-k_{v}\right)}{4kk_{v}}+\frac{A}{4kk_{v}}-\frac{1}{4k}.\label{eq:alternative coefficients}
\end{eqnarray}
which have some obvious problems. The result for $B_{1}$ does not
depend on $k_{v}$ and does not agree with the result for $B$ in
the collinear limit given in (\ref{eq:collinear coeff solutions}).
The result for $C$ is singular in the limit $k_{v}\rightarrow0$.
These problems are what necessitates the inclusion of the final term
in the assumed solution in Eq. (\ref{eq:special noncollinear solution}).

We now return to the solution of the constraints in Eqs (\ref{eq:special non-collinear 1})
- (\ref{eq:limiting conditions for C}) for the unknown coefficients
in the assumed solution in Eq. (\ref{eq:special noncollinear KG}).
The first constraint equation, Eq. (\ref{eq:special non-collinear 1}),
is equivalent to the lightcone condition (\ref{eq:light cone condition uv coords})
and leads to the result:
\begin{equation}
A=\frac{k_{x}^{2}+k_{y}^{2}}{4k_{v}},\label{eq:Asoln}
\end{equation}
which from Eq. (\ref{eq:limiting A}) has the collinear limit:
\begin{equation}
\lim_{\theta\rightarrow0}A=-\omega.\label{eq:Asolncollinear}
\end{equation}
Solving Eq. (\ref{eq:special non-collinear 2}) for $B_{2}$ gives
the result:
\begin{equation}
B_{2}=\frac{3A+k}{2k}-B_{1}.\label{eq:B2soln}
\end{equation}
Substituting this result into Eq. (\ref{eq:special non-collinear 3})
and solving for $B_{1}$ gives:
\begin{equation}
B_{1}=\frac{A}{4kB_{3}}-\frac{\left(3A+k\right)}{4k^{2}B_{3}}\left(\frac{A+2k}{k}-2B_{3}\right)+\frac{k_{v}\left(4Ck^{2}-3A\right)}{4k^{2}B_{3}}.\label{eq:equation for B1}
\end{equation}
If we set:
\begin{equation}
B_{3}=\frac{A+2k}{2k}\label{eq:B3soln}
\end{equation}
this becomes:
\begin{equation}
B_{1}=\frac{A}{2\left(A+2k\right)}+\frac{\left(3A-4Ck^{2}\right)k_{v}}{2k\left(A+2k\right)}.\label{eq:B1soln}
\end{equation}
This solution for $B_{3}$ ensures that $B_{1}$ has the correct collinear
limit. It can be seen that we have used constraint equations Eq. (\ref{eq:special non-collinear 3})
\& (\ref{eq:limiting B1}) to solve for $B_{3}$ and $B_{1}$. 

Now considering Eq. (\ref{eq:special non-collinear 4}), it can be
seen that the terms that are singular in the collinear limit can be
eliminated by requiring: 
\begin{equation}
\frac{B_{2}}{k_{v}}\left[3B_{3}\left(A+2k\right)-4k\left(B_{3}^{2}+B_{5}\right)\right]=0\label{eq:finiteness constraint}
\end{equation}
which is equivalent to the finiteness constraint in (\ref{eq:limiting conditions for C}).
This constraint can be solved by:
\begin{equation}
B_{5}=\frac{1}{2}B_{3}^{2}=\frac{\left(A+2k\right)^{2}}{8k^{2}}.\label{eq:B5soln}
\end{equation}
The remaining nonsingular part of Eq. (\ref{eq:special non-collinear 4})
is:
\begin{equation}
-3C\left(A+2k-k_{v}\right)-B_{2}\left(3B_{3}+4kB_{4}\right)+3\left(B_{1}+B_{2}\right)=0.\label{eq:-8}
\end{equation}
Using the results (\ref{eq:B2soln}) , (\ref{eq:B3soln}) and (\ref{eq:B1soln})
it can be shown that:
\begin{eqnarray}
C & = & \frac{AB_{3}\left(3B_{3}+4B_{4}k\right)}{2\left(A+2k\right)\left(3B_{3}\left(A+2k\right)+4B_{4}kk_{v}\right)}-\frac{B_{3}\left(3A+k\right)\left(3B_{3}+4B_{4}k-3\right)}{2k\left(3B_{3}\left(A+2k\right)+4B_{4}kk_{v}\right)}\nonumber \\
 &  & +\frac{3A\left(3B_{3}+4B_{4}k\right)k_{v}}{4k^{2}\left(3B_{3}\left(A+2k\right)+4B_{4}kk_{v}\right)}.\label{eq:C equation}
\end{eqnarray}
It can further be shown that if:
\begin{equation}
B_{4}=\frac{3(1-B_{3})}{4k}+\frac{3A(1-B_{3})k_{v}}{4\left(B_{3}\left(A+2k\right)\left(Ak-\left(A+2k\right)\left(3A+k\right)\right)-Akk_{v}\right)}\label{eq:B4soln}
\end{equation}
then the solution for $C$ will be:
\begin{equation}
C=\frac{A}{2(A+2k)^{2}}+\frac{3A\left(3B_{3}+4B_{4}k\right)k_{v}}{4k^{2}\left(3B_{3}(A+2k)+4B_{4}kk_{v}\right)}\label{eq:Csoln}
\end{equation}
which has the correct collinear limit as indicated in (\ref{eq:limiting conditions for C}).
It can be seen that we have used three constraints from (\ref{eq:special non-collinear 4})
and (\ref{eq:limiting conditions for C}) to determine the three coefficients
$B_{4}$, $B_{5}$ and $C$.

Finally from Eq. (\ref{eq:special non-collinear 5}) and Eq. (\ref{eq:B2soln})
it can be shown that:
\begin{equation}
D=\frac{9A^{2}+16Ak+3k^{2}}{8k^{2}}.\label{eq:Dsoln}
\end{equation}
One can see easily see from equations (\ref{eq:Asolncollinear}),
(\ref{eq:B2soln}), (\ref{eq:B3soln}), (\ref{eq:B1soln}), (\ref{eq:B5soln}),
(\ref{eq:B4soln}), (\ref{eq:Csoln}) and (\ref{eq:Dsoln}) that the
solutions for all coefficients are finite in the collinear limit ($k_{v}\rightarrow0$)
and that $B_{1}$ and $C$ go to the correct values in the collinear
limit. The value of $D$ was not constrained in the previous section
where we took the collinear limit of the KG equation before solving
it (see Eq. (\ref{eq:KG collinear})), but is constrained in this
case by taking the collinear limit after solving the KG equation in
Eq. (\ref{eq:special noncollinear KG}). The solution of the KG equation
in this special noncollinear case is given by substituting the solutions
for the coefficients from (\ref{eq:Asoln}), (\ref{eq:B2soln}), (\ref{eq:B3soln}),
(\ref{eq:B1soln}), (\ref{eq:B5soln}), (\ref{eq:B4soln}), (\ref{eq:Csoln})
and (\ref{eq:Dsoln}) into Eq. (\ref{eq:special noncollinear solution}).

It is useful to compare this result with the result of \cite{Jones - Scalar Production}
since they considered the same special case except with a plane gravitational
wave spacetime background. They solved Eq. (\ref{eq:Preston KG})
in the special case where $k_{x}=k_{y}$ so that they could find the
exact solution:
\begin{equation}
U(u)={\cal A}e^{\nicefrac{\lambda}{k}}e^{\frac{-\lambda}{k\left(1-h_{+}^{2}\,e^{2iku}\right)}}\left(1-h_{+}^{2}\,e^{2iku}\right)^{\frac{1}{2}\left(\frac{\lambda}{k}-1\right)}e^{-i\lambda u}\label{eq:Preston solution}
\end{equation}
which can be expanded in a series solution to give:
\begin{equation}
U(u)\approx e^{-i\lambda u}\left(1+\frac{\left(k-3\lambda\right)}{2k}h_{+}^{2}\,e^{2iku}+\frac{\left(9\lambda^{2}-16k\lambda+3k^{2}\right)}{8k^{2}}h_{+}^{4}\,e^{4iku}\right).\label{eq:Preston expansion}
\end{equation}
We can see that the coefficients of this expansion agree with the
expressions for $B_{1}$ in Eq. (\ref{eq:alternative coefficients})
and for $D$ in (\ref{eq:Dsoln}) with the substitution $A\rightarrow-\lambda$.
Since, as we discussed in the previous section, there is no constraint
on the collinear solution in their gravitational wave model there
is no problem with this result for $B_{1}$ in their analysis. Comparing
the expansion in Eq. (\ref{eq:Preston expansion}) with the solution
in Eq. (\ref{eq:special noncollinear solution}) we can see how the
solution for our gravitational wave model extends upon the results
of \cite{Jones - Scalar Production} for a plane gravitational wave.
We will leave discussion of the physical implications of our solution
until Section \ref{sec:Physical-Implications}.

\section{General Noncollinear Case\label{sec:General-Non-Collinear-Case}}

In this section we present the solution to Eq. (\ref{eq:our DE})
in the general case for a scalar wave travelling in an arbitrary direction
with arbitrary values of $k_{v}$, $k_{x}$ and $k_{y}$. We present
the assumed form of the solution and the solutions for the coefficients.
The constraint equations are quite long and are given in Appendix
A. The details of the solution process of these constraint equations
do not provide any new insights and are therefore omitted. 

The assumed form of the solution is:
\begin{eqnarray}
\Psi\left(u,x,y,v\right) & = & {\cal A}e^{iAu}e^{ik_{v}v}e^{ik_{x}x}e^{ik_{y}y}\left[1+\frac{4B_{1}\,h^{2}e^{2iku}}{\left(v-u\right)^{2}}+\frac{8i\,C\,h^{2}e^{2iku}}{\left(v-u\right)^{3}}+\frac{16\,D\,h^{4}e^{4iku}}{\left(v-u\right)^{4}}\right)\nonumber \\
 &  & \hspace{1.2in}+\frac{4\,B_{2}\,h^{2}\,e^{2iku}}{\left(v-u\right)^{2}}\left(1+\frac{2iB_{3}}{k_{v}\left(v-u\right)}+\frac{4\left(B_{4}k_{v}+B_{5}\right)}{k_{v}^{2}\left(v-u\right)^{2}}\right)^{-1}\nonumber \\
 &  & \hspace{1.2in}+\frac{2\,E\,h\,e^{iku}}{\left(v-u\right)}\left(1-\frac{2iE_{2}}{k_{v}\left(v-u\right)}+\frac{4E_{3}}{k_{v}^{2}\left(v-u\right)^{2}}-\frac{8iE_{4}}{k_{v}^{3}\left(v-u\right)^{3}}\right)^{-1}\nonumber \\
 &  & \hspace{1.2in}\left.+\frac{8\,F\,h^{3}\,e^{3iku}}{\left(v-u\right)^{3}}\left(1-\frac{2iF_{2}}{k_{v}\left(v-u\right)}\right)^{-1}\right].\label{eq:general solution}
\end{eqnarray}
Substituting this assumed solution into Eq. (\ref{eq:our DE}) and
following the procedure used in the previous section leads to the
following results for the coefficients:
\begin{eqnarray}
A=\frac{k_{x}^{2}+k_{y}^{2}}{4k_{v}}=\frac{\omega^{2}-k_{z}^{2}}{4\,k_{v}},\hspace{0.4in}E=-\frac{k_{x}^{2}-k_{y}^{2}}{2kk_{v}},\hspace{0.45in} &  & B_{2}=\frac{3A+k}{2k}+\frac{E^{2}}{2}-B_{1},\label{eq:A E B2 solution}\\
E_{3}=\frac{E\left(A+k-k_{v}\right)^{2}}{4k^{2}}+\frac{k_{v}}{2k},\hspace{1.6in} &  & E_{2}=-\frac{E\left(A+k-k_{v}\right)}{2k},\label{eq: E3 E2 solution}\\
D=\frac{9A^{2}+16Ak+3k^{2}}{8k^{2}}+\frac{E\left(3A\,E+5k\,E+2k\,F\right)}{8k},\hspace{0.2in} &  & F=\frac{E\left(9A+7k+k\,E^{2}\right)}{6k},\label{eq:D F solution}
\end{eqnarray}

\begin{eqnarray}
B_{3} & = & \frac{A+2k}{2k}-\frac{E^{3}\left(A+2k\right)\left(A+k-k_{v}\right)}{2\left(Ak-\left(A+2k\right)\left(3A+k+E^{2}k\right)\right)},\label{eq:B3 general solution}\\
B_{4} & = & \frac{3(1-B_{3})}{4k}+\frac{3A(1-B_{3})k_{v}}{4\left(B_{3}\left(A+2k\right)\left(Ak-\left(A+2k\right)\left(3A+k+E^{2}k\right)\right)-Akk_{v}\right)},\label{eq:B4 solution}\\
B_{5} & = & -B_{3}^{2}+\frac{3B_{3}\left(A+2k\right)}{4k}+\frac{E^{2}\left(\left(E+E^{2}\right)\left(A+k-k_{v}\right)^{2}+2kk_{v}\right)}{8B_{2}k^{2}},\label{eq:B5 general solution}\\
C & = & \frac{A}{2(A+2k)^{2}}+\frac{\left(3B_{3}+4B_{4}k\right)\left(3A+E^{2}k\right)k_{v}}{4k^{2}\left(3B_{3}(A+2k)+4B_{4}kk_{v}\right)},\label{eq: C general solution}\\
B_{1} & = & \frac{A}{2\left(A+2k\right)}+\frac{\left(3A+E^{2}k-4Ck^{2}\right)k_{v}}{4B_{3}k^{2}},\label{eq: B1 general solution}
\end{eqnarray}

\begin{eqnarray}
E_{4} & = & \frac{E\left(1+2E\right)\left(A+k-k_{v}\right)^{3}}{8k^{3}}+\frac{\left(1+3E\right)\left(A+k-k_{v}\right)k_{v}}{4k^{2}},\label{eq:E4 solution}\\
F_{2} & = & \frac{E\left(3A+4k-3k_{v}\right)}{6Fk}-\frac{E\left(A+k-k_{v}\right)}{2k}-\frac{\left(A+3k-k_{v}\right)}{2k}-\frac{E\left(B_{2}B_{3}-Ck_{v}\right)}{3F}\nonumber \\
 &  & +\frac{E^{2}\left(3A+k+E^{2}k\right)\left(A+k-k_{v}\right)}{12Fk^{2}}.\label{eq:F2 solution}
\end{eqnarray}
These expressions are quite long and hard to interpret. We can, however,
make a few important statements about these results:
\begin{enumerate}
\item These results determine the solution, up to order $1/z^{4}$, in the
general case of a scalar plane wave travelling in an arbitrary direction
in a spacetime background of a gravitational wave with amplitude that
decays with distance.
\item The coefficient $E$ is determined by the $x$ \& $y$ components
of the direction of propagation of the scalar plane wave which are
perpendicular to direction of propagation of the gravitational wave.
$E$ vanishes in the case where $k_{x}=\pm k_{y}$. In the limit where
$E\rightarrow0$, the expressions for $B_{1}\mbox{ to }B_{5},C$ and
$D$ agree with the results from the previous section. As well, $F$
also vanishes when $E$ vanishes and Eq. (\ref{eq:general solution})
agrees with the solution in Eq. (\ref{eq:special noncollinear solution})
for the special collinear case in the limit where $E=0.$
\item The solutions for $E,$ $E_{2}$ to $E_{4}$, $F$ and $F_{2}$ are
all finite in the collinear limit $\left(k_{v}\rightarrow0\right)$.
As a consequence, it can be seen that the terms in Eq. (\ref{eq:general solution})
involving these coefficients all vanish in the collinear limit, ($k_{v}\rightarrow0$).
In addition, since the coefficient $E$ disappears from the solution
in the collinear limit we can take $E=0$ in the solutions for the
other variables. Finally, as also discussed in the previous section,
the term in Eq. (\ref{eq:general solution}) involving $B_{2}$ to
$B_{5}$ vanishes in the collinear limit $\left(k_{v}\rightarrow0\right)$.
As a result it can easily be seen that the collinear limit of the
general solution in Eq. (\ref{eq:general solution}) agrees with the
collinear solution in Eq. (\ref{eq:collinear assumed solution}).
It is important to stress that approximating the solution in Eq. (\ref{eq:general solution})
as a series expansion in $\nicefrac{1}{z}=\nicefrac{2}{\left(v-u\right)}$
would result in a solution which does not have the correct collinear
limit.
\end{enumerate}
We will discuss the physical implications of this solution in Section
\ref{sec:Physical-Implications}. In the following sections we will
calculate four-currents from this general solution and discuss the
physical implications of this solution in the general case and in
the collinear case.

Before moving on, we compare this result to the results of \cite{Dasgupta - Scalar Fermion}.
The lowest order perturbation for the wave function in Eq. (\ref{eq:general solution})
is:
\begin{eqnarray}
{\cal A}e^{iku}e^{ik_{v}v}e^{ik_{x}x}e^{ik_{y}y}\frac{2\,E\,h\,}{\left(v-u\right)} & = & {\cal A}e^{iku}e^{ik_{v}v}e^{ik_{x}x}e^{ik_{y}y}\frac{h}{z}\frac{k_{y}^{2}-k_{x}^{2}}{2kk_{v}}.\label{eq:-9}
\end{eqnarray}
For simplicity we can consider the special case where $k_{y}=0$ and
$k_{x}=\omega\sin\theta$:

\begin{equation}
{\cal A}e^{iku}e^{ik_{v}v}e^{ik_{x}x}\frac{h}{z}\frac{k_{y}^{2}-k_{x}^{2}}{2kk_{v}}={\cal A}e^{iku}e^{ik_{v}v}e^{ik_{x}x}\frac{h}{z}\frac{\omega^{2}\sin^{2}\theta}{\omega k\left(1-\cos\theta\right)}\propto\frac{h}{z}{\cal A}\left(\frac{\omega}{\omega_{g}}\right)\label{eq:our lowest order}
\end{equation}
where $\nicefrac{h}{z}$ is the gravitational wave strain, $\omega$
is the angular frequency of the scalar wave and $\omega_{g}=kc$ is
the angular frequency of the gravitational wave after converting to
SI units. This result is equivalent to Eq. (39) in \cite{Dasgupta - Scalar Fermion}:
\begin{equation}
\Delta\phi\approx hA_{0}\left(\frac{\omega_{0}}{\omega}\right)\label{eq:Morales lowest order}
\end{equation}
where, in their notation, $h$ is the gravitational wave strain, $\omega_{0}$
is the angular frequency of the scalar wave, $\omega$ is the angular
frequency of the gravitational wave and $A_{0}$ is the coefficient
of the unperturbed wavefunction. Therefore, in the general noncollinear
case, our power series solution of the KG equation in the spacetime
background of a gravitational wave with spatially dependent amplitude
is consistent, up to first order in the gravitational strain, with
the solution of \cite{Dasgupta - Scalar Fermion} for a plane gravitational
wave background.

\section{Four-Current Components\label{sec:Four-Current}}

In order to discuss the physical implications of our solutions it
is useful to calculate the scalar four-current associated with this
solution. It can easily be seen from Eq. (\ref{eq:general solution})
that the components of the scalar four-current in the $x$ and $y$
directions:
\begin{eqnarray}
j_{x} & = & -i\left(\Psi^{*}\partial_{x}\Psi-\Psi\,\partial_{x}\Psi^{*}\right)=2{\cal A}^{2}k_{x},\label{eq:jx}\\
j_{y} & = & -i\left(\Psi^{*}\partial_{y}\Psi-\Psi\,\partial_{y}\Psi^{*}\right)=2{\cal A}^{2}k_{y},\label{eq:jy}
\end{eqnarray}
are independent of $h$ and are unaffected by the presence of the
gravitational wave. The four-current components in the $z$ and $t$
directions are affected by the gravitational wave:
\begin{eqnarray}
j_{t} & = & -i\left(\Psi^{*}\partial_{t}\Psi-\Psi\,\partial_{t}\Psi^{*}\right)=-2{\cal A}^{2}\omega+{\cal O}\left(h\right),\label{eq:jt}\\
j_{z} & = & -i\left(\Psi^{*}\partial_{z}\Psi-\Psi\,\partial_{z}\Psi^{*}\right)=2{\cal A}^{2}\omega\cos\theta+{\cal O}\left(h\right).\label{eq:jz}
\end{eqnarray}
The full expressions for $j_{t}$ and $j_{z}$ are given in Appendix
B. In order to determine these expressions all terms of the form $e^{iku},$
$e^{2iku}$ , $e^{3iku}$ and $e^{4iku}$ in $\Psi$ were expanded
in terms of sine and cosine functions. The $e^{iAu}e^{ik_{v}v}e^{ik_{x}x}e^{ik_{x}y}$
terms were left in exponential form as they cancel out in the calculations
of $j_{z}$ and $j_{t}$. After taking derivatives, $\Psi^{*}\partial_{\mu}\Psi$
and $\Psi\partial_{\mu}\Psi^{*}$ contain terms including all two
factor products of $\cos\left(ku\right)$, $\cos\left(2ku\right)$,
$\cos\left(3ku\right)$, $\cos\left(4ku\right)$, $\sin\left(ku\right)$,
$\sin\left(2ku\right)$, $\sin\left(3ku\right)$ and $\sin\left(4ku\right)$.
Using trigonometric identities, these expressions were simplified
to terms proportional to single factors of $\cos\left(ku\right)$,
$\cos\left(2ku\right)$, $\cos\left(3ku\right)$, $\cos\left(4ku\right)$,
$\sin\left(ku\right)$, $\sin\left(2ku\right)$, $\sin\left(3ku\right)$
and $\sin\left(4ku\right)$ and non-sinusoidal terms. The variation
of $j_{z}$ with time is a possible observable signal of this interaction
which we will discuss in Section \ref{sec:Physical-Implications}.
Another possible effect that we will discuss in Section \ref{sec:Physical-Implications}
involves the cumulative effect of the gravitational wave. The relevant
quantity for the latter effect is the time-averaged current over a
period of the gravitational wave:
\begin{equation}
\left\langle j_{z}\right\rangle =\frac{1}{\nicefrac{2\pi}{k}}\int_{0}^{\nicefrac{2\pi}{k}}j_{z}\,dt.\label{eq:ave jz}
\end{equation}
Time averaging eliminates oscillating terms in $j_{z}$ and only includes
terms which are independent of $t$:
\begin{eqnarray}
\left\langle j_{z}\right\rangle  & = & 2{\cal A}^{2}\omega\cos\theta-{\cal A}^{2}\,\frac{2E^{2}h^{2}k_{v}^{7}z^{4}\left(3E_{4}k_{v}^{2}z^{2}+E_{2}k_{v}^{4}z^{4}+E_{3}\left(E_{4}-E_{2}k_{v}^{2}z^{2}\right)\right)}{\left(k_{v}^{2}z^{2}\left(E_{3}+k_{v}^{2}z^{2}\right)^{2}+\left(E_{4}+E_{2}k_{v}^{2}z^{2}\right)^{2}\right)^{2}}\nonumber \\
 &  & -{\cal A}^{2}\,\frac{2E^{2}h^{2}k_{v}^{6}z^{4}\left(k-\omega\cos\theta\right)}{k_{v}^{2}z^{2}\left(E_{3}+k_{v}^{2}z^{2}\right)^{2}+\left(E_{4}+E_{2}k_{v}^{2}z^{2}\right)^{2}}\nonumber \\
 &  & -2{\cal A}^{2}\left(2k-\omega\cos\theta\right)\left(\frac{B_{1}^{2}h^{4}}{z^{4}}+\frac{B_{2}^{2}h^{4}k_{v}^{2}\left(B_{2}k_{v}^{2}z^{2}+2B_{1}\left(B_{5}+B_{4}k_{v}+k_{v}^{2}z^{2}\right)\right)}{z^{2}\left(B_{3}^{2}k_{v}^{2}z^{2}+\left(B_{5}+B_{4}k_{v}+k_{v}^{2}z^{2}\right)^{2}\right)}\right).\label{eq:general ave current}
\end{eqnarray}
This expression seems like it could be simplified in terms of a power
series in $1/z$ and this expansion is valid as long as we are not
too close to the collinear limit. We will discuss the range of validity
shortly, after first using this expansion to compare our result with
previous results in the literature.

The series expansion of $\left\langle j_{z}\right\rangle $ up to
order $\left(1/z\right)^{4}$ is:
\begin{eqnarray}
\left\langle j_{z}\right\rangle _{approx} & = & 2{\cal A}^{2}\omega\cos\theta-{\cal A}^{2}\,\frac{2E^{2}h^{2}\left(k-\omega\cos\theta\right)}{z^{2}}+{\cal A}^{2}\,\frac{E^{2}h^{2}\left(2+E\right)\left(k-\omega\cos\theta\right)}{kk_{v}z^{4}}\nonumber \\
 &  & -{\cal A}^{2}\,\frac{h^{4}\left(3A+k\right)^{2}\left(2k-\omega\cos\theta\right)}{2k^{2}z^{4}}+{\cal A}^{2}\,\frac{E^{3}h^{2}\left(2+E\right)\left(k-\omega\cos\theta\right)^{3}}{2k^{2}k_{v}^{2}z^{4}}.\label{eq:approx ave curr=002026}
\end{eqnarray}
In order to compare with the results of \cite{Jones - Scalar Production},
we take the special case where $k_{x}=\pm k_{y}$ which implies $E=0$
and use the relation:
\begin{equation}
\left(A-k_{v}\right)=-\text{\ensuremath{\omega\cos\theta}}\label{eq:-17}
\end{equation}
which can be easily verified using the relations in (\ref{eq:lightcone momentum components})
where $k_{u}=A$. In this case we have that:
\begin{eqnarray}
\lim_{E\rightarrow0}\left\langle j_{z}\right\rangle _{approx} & \hspace{-0.4cm}= & \mbox{\hspace{-0.4cm}}-2{\cal A}^{2}\left(A-k_{v}\right)-{\cal A}^{2}\,\frac{h^{4}\left(3A+k\right)^{2}\left(A+2k-k_{v}\right)}{2k^{2}z^{4}}\nonumber \\
 & \mbox{\hspace{-0.4cm}}= & \mbox{\hspace{-0.4cm}}-2A{\cal A}^{2}-{\cal A}^{2}\frac{h^{4}}{z^{4}}\left(\frac{9A^{3}}{2k^{2}}+\frac{12A^{2}}{k}+\frac{13A}{2}+k\right)+{\cal A}^{2}k_{v}\left(2+\frac{h^{4}\left(3A+k\right)^{2}}{2k^{2}z^{4}}\right)\negthickspace.\label{eq:special approx ave jz}
\end{eqnarray}
We can similarly take the time average of the time component of the
four-current, $j_{t}$ , over one period of the gravitational wave,
series expand this result and take the $E\rightarrow0$ limit to give:
\begin{eqnarray}
\lim_{E\rightarrow0}\left\langle j_{t}\right\rangle _{approx} & \hspace{-0.4cm}= & \mbox{\hspace{-0.2cm}}2{\cal A}^{2}\left(A+k_{v}\right)+{\cal A}^{2}\,\frac{h^{4}\left(3A+k\right)^{2}\left(A+2k+k_{v}\right)}{2k^{2}z^{4}}\nonumber \\
 & \mbox{\hspace{-0.4cm}}= & \mbox{\hspace{-0.2cm}}2A{\cal A}^{2}+{\cal A}^{2}\frac{h^{4}}{z^{4}}\left(\frac{9A^{3}}{2k^{2}}+\frac{12A^{2}}{k}+\frac{13A}{2}+k\right)+{\cal A}^{2}k_{v}\left(2+\frac{h^{4}\left(3A+k\right)^{2}}{2k^{2}z^{4}}\right)\label{eq:special approx ave jt}
\end{eqnarray}
and then calculate the four-current component in the $u$ direction:
\begin{eqnarray}
\lim_{E\rightarrow0}\left\langle j_{u}\right\rangle _{approx} & = & \lim_{E\rightarrow0}\frac{1}{2}\left(\left\langle j_{t}\right\rangle _{approx}-\left\langle j_{z}\right\rangle _{approx}\right)\nonumber \\
 & = & 2A{\cal A}^{2}+{\cal A}^{2}\frac{h^{4}}{z^{4}}\left(\frac{9A^{3}}{2k^{2}}+\frac{12A^{2}}{k}+\frac{13A}{2}+k\right)\label{eq:-10}
\end{eqnarray}
This result is the same as Eq. (19) in \cite{Jones - Scalar Production}:
\begin{equation}
\left\langle j_{u}\right\rangle =-2\lambda A^{2}-A^{2}\frac{h^{4}}{z^{4}}\left(\frac{9\lambda^{3}}{2k^{2}}-\frac{12\lambda^{2}}{k}+\frac{13\lambda}{2}-k\right).\label{eq:preston ave current}
\end{equation}
after converting to their notation with the substitutions ${\cal A}\rightarrow A$
and $A\rightarrow-\lambda$. Therefore away from the collinear limit,
where this approximation is valid, our result for a gravitational
wave with amplitude that decreases with distance agrees with the result
of \cite{Jones - Scalar Production} for a plane gravitational wave. 

Now consider the collinear limit of the series expansion of $\left\langle j_{z}\right\rangle $
in Eq. (\ref{eq:approx ave curr=002026}):
\begin{equation}
\lim_{k_{v},E,F\rightarrow0}\left\langle j_{z}\right\rangle _{approx}=2\omega{\cal A}^{2}+{\cal A}\,\frac{h^{4}\left(3\omega-k\right)^{2}\left(\omega-2k\right)}{2k^{2}z^{4}}.\label{eq:jz approx collinear}
\end{equation}
The collinear limit $\left(k_{v}\rightarrow0\right)$ of Eq. (\ref{eq:general ave current})
can be shown to be:
\begin{eqnarray}
\left\langle j_{z}\right\rangle _{collinear} & = & 2{\cal A}^{2}+{\cal A}^{2}\frac{\omega^{2}h^{4}}{2\left(\omega-2k\right)z^{4}}\label{eq:collinear ave current}
\end{eqnarray}
where we have used (\ref{eq:limiting B1}) and (\ref{eq:Asolncollinear}).
The collinear limit in Eq. (\ref{eq:collinear ave current}) is a
quite different from Eq. (\ref{eq:jz approx collinear}). The form
of $\left\langle j_{z}\right\rangle _{collinear}$ in Eq. (\ref{eq:collinear ave current})
is interesting because it has resonant behaviour for $\omega\approx2k$
which is not present in (\ref{eq:approx ave curr=002026}) or (\ref{eq:preston ave current}).
We will discuss this resonance effect further in Section \ref{sec:Physical-Implications}.
The collinear limit of the expansion in Eq. (\ref{eq:approx ave curr=002026})
is incorrect for the same reason that the collinear limit of the series
expansions of the last term in Eq. (\ref{eq:special noncollinear solution})
and the last three terms in  Eq. (\ref{eq:general solution}) would
lead to incorrect results. The issue in all these cases is repeated
occurrences of the term $k_{v}z$ which goes to $\infty$ when $z\rightarrow\infty$
but goes to $0$ when $k_{v}\rightarrow0$. We will discuss this issue
in detail for $\left\langle j_{z}\right\rangle $, but most of the
arguments also apply to the expansion of terms in Eq. (\ref{eq:special noncollinear solution})
and Eq. (\ref{eq:general solution}).

Doing the series expansion of (\ref{eq:general ave current}) to get
(\ref{eq:approx ave curr=002026}) involves, for very large $z$,
grouping the contributions of different powers of $k_{v}z$ according
to the power of $1/z$. This is valid for most of the range of possible
values of $k_{v}$. However, as the collinear limit is approached,
when $k_{v}z\sim1$, all of these different powers of $k_{v}z$ are
of approximately the same size and all of these terms are important.
Therefore it is necessary to use Eq. (\ref{eq:general ave current})
for the time-averaged current when $k_{v}z\sim1$ and this is the
only expression that has the correct form near the collinear limit
when $k_{v}z\rightarrow0$. On the other hand Eq. (\ref{eq:approx ave curr=002026})
is a good approximation for $\left|k_{v}\right|z\gnapprox1$.

We will end this section with a discussion of the range of $\theta$
over which the approximate expression in Eq. (\ref{eq:approx ave curr=002026})
is valid. We can use Eq. (\ref{eq:lightcone momentum components})
and a trigonometric identity to write the restriction on $\theta$
as:
\begin{eqnarray}
\left|k_{v}\right|z=\omega z\frac{\left(1-\cos\theta\right)}{2}\gnapprox1 & \Rightarrow & \sin\left(\frac{\theta}{2}\right)\gnapprox\sqrt{\frac{c}{\omega z}}\label{eq:collinear angle limits}
\end{eqnarray}
where we have converted to SI units in the second expression. For
the closest observed gravitational wave source \cite{Third Run Second Half,Third Run First Half,First and Second Runs},
$z\approx35Mpc\approx1\times10^{24}\,m$ and for a scalar plane wave
with $\omega>10^{-6}\,s^{-1}$:
\begin{eqnarray}
\sqrt{\frac{c}{\omega z}} & < & 10^{-5}\label{eq: numerical angle limit}
\end{eqnarray}
which means that for realistic distances and frequencies, the approximate
solution in Eq. (\ref{eq:approx ave curr=002026}) is valid even for
very small angles where $\theta\gnapprox0.001^{\circ}$. The extremely
narrow range for which the full solution in Eq. (\ref{eq:general ave current})
is necessary is a direct result of the large distances to currently
known gravitational wave sources and is not a fundamental property
of this analysis. For instance, for an observer at a distance of 1
light year from the source of the same gravitational wave the approximate
expression is only valid for $\theta\gnapprox20^{\circ}$.

This discussion about the range of values of $\theta$ over which
the approximate expression for $\left\langle j_{z}\right\rangle $
is valid would also be applicable to the solutions in Eq'ns (\ref{eq:special noncollinear solution})
and (\ref{eq:general solution}). Series expansions of the last term
in Eq. (\ref{eq:special noncollinear solution}) and the last three
terms in Eq. (\ref{eq:general solution}) would give good approximations
to the solutions except for scalar waves which are very close to collinear
$\left(\theta\lesssim\sqrt{\frac{4c}{\omega z}}\right)$ with the
gravitational wave.

\section{No Effect in a Vacuum\label{sec:No-Particle-Production}}

The solutions to the KG equations that we have discussed all assume
the presence of a scalar plane wave that exists in the absence of
the gravitational wave. Taking the limit $h\rightarrow0$ of the solutions
in (\ref{eq:collinear solution}), (\ref{eq:special noncollinear solution}),
and (\ref{eq:general solution}) give scalar plane waves in a flat
spacetime. We cannot, however, ``turn off'' the scalar plane wave
to obtain a scalar excitation that is entirely produced by the gravitational
wave. 

The collinear limit ($k_{v},E,F\rightarrow0$) of the general solution
in Eq. (\ref{eq:general solution}) produces the result in Eq. (\ref{eq:collinear solution}).
This solution and contingent results are valid for arbitrarily small
values of $A=-\omega$ which corresponds to arbitrarily long wavelength
scalar waves. If you consider taking $\omega=0$, however, this leads
to the result:
\begin{equation}
\Psi\left(u,v\right)={\cal A}\left(1+\frac{4B\,h^{2}e^{2iku}}{\left(v-u\right)^{2}}+\frac{8i\,C\,h^{2}e^{2iku}}{\left(v-u\right)^{3}}+\frac{16\,D\,h^{4}e^{4iku}}{\left(v-u\right)^{4}}\right),\label{eq:long wavelength limit}
\end{equation}
which is a waveform that oscillates about $\Psi={\cal A}$. In fact,
if you were to eliminate the gravitational wave ($h\rightarrow0$)
this leads to the constant solution $\Psi\left(u,v\right)={\cal A}$.
Other authors have previously suggested that this indicates that the
gravitational wave leads to a nonzero vacuum expectation value of
the massless scalar field via a Higgs-like mechanism \cite{Jones - Scalar Production,Jones - Vacuum Exp}.
It is conceivable that this could happen, but to justify this conclusion
in our analysis we need to check if there is an alternative form of
the solution which avoids the nonzero vacuum expectation value. 

The collinear solution in Eq. (\ref{eq:collinear solution}) involves
a scalar wave moving in the same direction as the gravitational wave.
It is not clear, however, that the effects of a gravitational wave
on the scalar field would necessarily only result in a collinear scalar
wave moving in the positive $z-$direction. We need to also consider
a scalar wave that moves in the $-z$ direction, or anti-collinear
to the gravitational wave. We can take the anti-collinear limit by
taking $A,\,k_{x},\,k_{y}\rightarrow0$ and $k_{v}\rightarrow-\omega$
which leads to $E\rightarrow0$, $F\rightarrow0$, $B_{1}\rightarrow0,$
$B_{3}\rightarrow1,$ $B_{4}\rightarrow0,$ $B_{5}\rightarrow\frac{1}{2}$,
$B_{2}\rightarrow\frac{1}{2}$, $C\rightarrow0$ and $D\rightarrow\frac{3}{8}$.
In this case, the general solution becomes:
\begin{eqnarray}
\Psi_{anticollinear}\left(u,v\right) & = & {\cal A}e^{-i\omega v}\left(1+\frac{6\,h^{4}e^{4iku}}{\left(v-u\right)^{4}}+\frac{2\,h^{2}e^{2iku}}{\left(v-u\right)^{2}}\left(1-\frac{2i}{\omega\left(v-u\right)}+\frac{2}{\omega{}^{2}\left(v-u\right)^{2}}\right)^{-1}\right)\nonumber \\
 & = & {\cal A}e^{-i\omega t}e^{-i\omega z}\left(1+\frac{3\,h^{4}e^{4iku}}{8z^{4}}+\frac{h^{2}e^{2iku}}{2z^{2}}\left(1-\frac{i}{\omega z}+\frac{1}{2\omega{}^{2}z^{2}}\right)^{-1}\right),\label{eq:anticollinear solution}
\end{eqnarray}
which can be compared to the collinear solution which can be written
in the form:
\begin{eqnarray}
\Psi_{collinear}\left(u,z\right) & = & {\cal A}e^{-i\omega t}e^{i\omega z}\left(1-\frac{\omega h^{2}e^{2iku}}{2\left(2k-\omega\right)z^{2}}-\frac{i\omega h^{2}\,e^{2iku}}{2\left(2k-\omega\right)^{2}z^{3}}+\frac{h^{4}e^{4iku}\left(9\omega^{2}-16\omega k+3k^{2}\right)}{8k^{2}z^{4}}\right).\label{eq:collinear solution-2}
\end{eqnarray}
These solutions are waves travelling in the $-z$ and $+z$ directions
respectively and are analogous to the travelling wave solutions $e^{-ik\left(x+t\right)}$
and $e^{ik\left(x-t\right)}$ in one dimensional QM. We are looking
for a solution which involves waves travelling in both directions
analogous to the $\sin\left(kx\right)$ or $\cos\left(kx\right)$
solutions in one dimensional QM. Since the KG equations in (\ref{eq:Full KG})
and (\ref{eq:KG collinear}) are linear we can consider the following
linear combination of the collinear and anti-collinear solutions:
\begin{eqnarray}
\Psi_{-}\left(u,t,z\right) & = & \Psi_{collinear}\left(u,z\right)-\Psi_{anticollinear}\left(u,z\right)\nonumber \\
 & = & {\cal A}e^{-i\omega t}\left(e^{i\omega z}-e^{-i\omega z}\right)\nonumber \\
 &  & +{\cal A}e^{-i\omega t}e^{i\omega z}\left(-\frac{\omega h^{2}e^{2iku}}{2\left(2k-\omega\right)z^{2}}-\frac{i\omega h^{2}\,e^{2iku}}{2\left(2k-\omega\right)^{2}z^{3}}+\frac{h^{4}e^{4iku}\left(9\omega^{2}-16\omega k+3k^{2}\right)}{8k^{2}z^{4}}\right)\nonumber \\
 &  & -{\cal A}e^{-i\omega t}e^{-i\omega z}\left(\frac{3\,h^{4}e^{4iku}}{8z^{4}}+\frac{h^{2}e^{2iku}}{2z^{2}}\left(1-\frac{i}{\omega z}+\frac{1}{2\omega{}^{2}z^{2}}\right)^{-1}\right)\nonumber \\
 & = & 2i{\cal A}\sin\left(\omega z\right)e^{-i\omega t}+\nonumber \\
 &  & -{\cal A}e^{-i\omega t}e^{i\omega z}\left(\frac{\omega h^{2}e^{2iku}}{2\left(2k-\omega\right)z^{2}}+\frac{i\omega h^{2}\,e^{2iku}}{2\left(2k-\omega\right)^{2}z^{3}}-\frac{h^{4}e^{4iku}\left(9\omega^{2}-16\omega k+3k^{2}\right)}{8k^{2}z^{4}}\right)\nonumber \\
 &  & -{\cal A}e^{-i\omega t}e^{-i\omega z}\left(\frac{3\,h^{4}e^{4iku}}{8z^{4}}+\frac{h^{2}e^{2iku}}{2z^{2}}\left(1-\frac{i}{\omega z}+\frac{1}{2\omega{}^{2}z^{2}}\right)^{-1}\right).\label{eq:sine solution}
\end{eqnarray}
This solution oscillates about zero which eliminates any need for
a Higgs-like mechanism to explain a nonzero vacuum expectation value.
Note that even though we have written the first term seperately from
the $h$ dependent terms to make this point clear, this term is not
independent of the other terms as they are all proportional to the
overall amplitude ${\cal A}$. If we take ${\cal A}$ to zero the
whole wavefunction vanishes. This is consistent with \cite{Dasgupta - Scalar Fermion}
where the perturbations they found were dependent on the solutions
to the KG equation in flat space, $\psi_{0}$, and the perturbations
vanish for the trivial solution $\psi_{0}=0.$\footnote{See Eq'ns (17), (31) and (32) in \cite{Dasgupta - Scalar Fermion}}.

If we consider the long wavelength limit as a way to effectively ``turn
off'' the independent scalar field, as done in\emph{ }\cite{Jones - Scalar Production},
we get: 
\begin{eqnarray}
\lim_{\omega\rightarrow0}\Psi_{-}\left(u,t,z\right) & = & \lim_{\omega\rightarrow0}\left[{\cal A}e^{-i\omega t}\left(\frac{h^{4}e^{4iku}\left(3k^{2}\right)}{8k^{2}z^{4}}\right)-{\cal A}e^{-i\omega t}\left(\frac{3\,h^{4}e^{4iku}}{8z^{4}}\right)\right]\nonumber \\
 & = & \lim_{\omega\rightarrow0}{\cal A}e^{-i\omega t}\left(\frac{3\,h^{4}e^{4iku}}{8z^{4}}-\frac{3\,h^{4}e^{4iku}}{8z^{4}}\right)=0.\label{eq:long wavelength limit of sine solution}
\end{eqnarray}
which shows that the solution vanishes in this limit. Therefore the
solution of the form $\Psi_{-}\left(u,t,z\right)$ is a solution of
the collinear KG equation which does not require any new physics to
explain it. This solution, however, does not allow us to consider
a scalar waveform that is produced entirely by the gravitational wave.
The conclusion we draw from this observation is that a gravitational
wave will enhance an existing scalar wave but will not produce a scalar
waveform in a vacuum. This conclusion is different from the results
of \cite{Jones - Scalar Production} but consistent with the results
of \cite{Dasgupta - Scalar Fermion}, both of which involved a plane
gravitational wave.

The physical implication of this conclusion is that gravitational
waves will not produce scalar particles in a vacuum, which is consistent
with the conclusions of \cite{Gibbons}. Gravitational waves can only
produce scalar particles by interacting with an independently existing
scalar plane wave as we will discuss in the next section. 

\section{Physical Implications\label{sec:Physical-Implications}}

We will now discuss three different types of physical implications
of the interaction of the gravitational wave with scalar plane waves.
In all these cases we assume that the scalar plane wave is part of
some existing background field analogous to the Cosmic Microwave Background
(CMB) or the Cosmic Neutrino Background (C$\nu B$) \cite{Steigman}.

\subsection{Sinusoidal Variation of $j_{z}$ in Time\label{subsec:Sinusoidal-Variation-of}}

The full expression for the time-dependent component of the scalar
four-current, $j_{z}$, up to order $\nicefrac{1}{z^{4}}$, is given
in Eq. (\ref{eq:time dependent jz}) in Appendix B. Here we only consider
terms of $j_{z}$ up to first order in $\nicefrac{1}{z}$:
\begin{equation}
j_{z}(u,v)\approx2{\cal A}^{2}\omega\cos\theta+2{\cal A}^{2}{\cal E}\left(k-2\omega\cos\theta\right)\cos\left(ku\right)\label{eq:sinusoidal variation}
\end{equation}
where:
\begin{equation}
{\cal E}=\frac{E\,hk_{v}^{4}z^{3}\left(E_{3}+k_{v}^{2}z^{2}\right)}{k_{v}^{2}z^{2}\left(E_{3}+k_{v}^{2}z^{2}\right)^{2}+\left(E_{4}+E_{2}k_{v}^{2}z^{2}\right)^{2}}.\label{eq:cal E definition}
\end{equation}
It can easily be seen that the oscillating term will vanish in the
collinear limit, $\left(k_{v}\rightarrow0\right)$. For noncollinear
contributions when $\left|k_{v}\right|z\gnapprox1$ or $\sin\left(\frac{\theta}{2}\right)\gnapprox\sqrt{\frac{c}{\omega z}}$
a series expansion in $1/z$ gives the approximate result:
\begin{eqnarray}
j_{z}(u,v) & \approx & 2\frac{\omega}{c}{\cal A}^{2}\cos\theta+2E{\cal A}^{2}\frac{h}{cz}\left(2\omega\cos\theta-\omega_{g}\right)\cos\left(ku\right)\nonumber \\
 & = & 2\frac{\omega}{c}{\cal A}^{2}\cos\theta+\frac{k_{y}^{2}-k_{x}^{2}}{kk_{v}}{\cal A}^{2}\frac{h}{cz}\left(2\omega\cos\theta-\omega_{g}\right)\cos\left(ku\right)\label{eq:first order time dep. jz}
\end{eqnarray}
where we have converted to SI units and $\omega_{g}=kc$ is the angular
frequency of the gravitational wave. The high frequency limit of current
earth-based gravitational wave detection experiments is of the order
of $f_{g}\sim1\,kHz$ \cite{Third Run Second Half}. The low frequency
limit of earth-based radio telecopes is roughly of the order of $f\sim10\,MHz.$
Therefore, we can assume that $2\omega\cos\theta\gg\omega_{g}$ except
for $\theta\approx\frac{\pi}{2}$. As we have discussed previously,
for gravitational waves with currently observed parameters the range
of $\theta$ over which the series expansion would not be valid is
very small. Therefore, for an estimate of the maximum magnitude of
oscillation we can use $\sqrt{\frac{4c}{\omega z}}\ll\theta\ll1$.
In this case, using (\ref{eq:lightcone momentum components}) and
(\ref{eq:limiting E coord. sing.}), the perturbation of the current
from that in the absence of a gravitational wave can be shown to be:
\begin{eqnarray}
\Delta j_{z,1}(u,v) & \approx & {\cal A}^{2}\frac{4h\omega}{z\omega_{g}}\left(2\omega\cos\theta\right)\cos^{2}\left(\nicefrac{\theta}{2}\right)\cos\left(2\phi\right)\cos\left[k\left(t-z\right)\right].\label{eq:sinusoidal perturbation}
\end{eqnarray}
For fixed values of $\phi,\,\theta$ and $z$ this leads to a sinusoidal
oscillation in time of the scalar current in the $z-$direction with
frequency equal to the frequency of the gravitational wave. This oscillation
in the current is essentially equivalent to the variation in the amplitude
of the wavefunction described in \cite{Dasgupta - Scalar Fermion}.
In this case the maximum amplitude of oscillation in $j_{z}$ relative
to the magnitude of $j_{z}$ in the absence of the gravitational wave
is given by:
\begin{equation}
\frac{\Delta j_{z,1}(u,v)}{j_{z,0}(u,v)}\equiv\frac{\Delta j_{z,1}(u,v)}{2{\cal A}^{2}\omega\cos\theta}\approx\frac{4h\omega}{z\omega_{g}}.\label{eq:}
\end{equation}
In order to obtain a numerical estimate of the relative amplitude
we need to make some further assumptions. The largest amplitude gravitational
waves detected have gravitational strains at earth of order $\nicefrac{h}{z}\sim10^{-21}$.
The low frequency limit of projected space-based gravitational wave
observatories is of the order of order $f_{g}\sim10^{-4}s^{-1}$ \cite{LISA}.
For scalar waves with $f\approx160\,GHz$, which is roughly the peak
frequency of the cosmic microwave background, the relative amplitude
of the variation in current would be approximately: 
\begin{equation}
\frac{\Delta j_{z,1}(u,v)}{j_{z,0}(u,v)}\approx\frac{4hf}{zf_{g}}\approx6\times10^{-6}\label{eq:-1}
\end{equation}
which should be a detectable variation. The variation in the current
predicted here is of the same order as the variation in the amplitude
of $\Psi$ discussed in \cite{Dasgupta - Scalar Fermion} which in
our notation\footnote{In the notation of \cite{Dasgupta - Scalar Fermion}, $h$ is the
dimensionless local gravitational strain, but in our notation the
dimensionless local strain is given by $h/z$. } would be:
\begin{equation}
\frac{h}{z}\frac{\omega}{\omega_{g}}=\frac{hf}{zf_{g}}\approx1.5\times10^{-6}.\label{eq:-2}
\end{equation}
Therefore, to first order in $\nicefrac{1}{z}$, the physical implications
from a gravitational wave that decreases with distance are consistent
with the results for a plane gravitational wave. 

We will briefly consider the variation of the term in $j_{z}$ involving
$B_{1}$ since it exhibits a resonance phenomenon which might make
it relevant. In the collinear limit $\left(\theta=0,k_{v}=0\right)$,
the second order perturbation to the current is:
\begin{eqnarray}
\Delta j_{z,2}(u,v) & \approx & -4{\cal A}^{2}{\cal C}\left(k-\omega\cos\theta\right)\cos\left(2ku\right)=-\frac{4B_{1}h^{2}}{z^{2}}{\cal A}^{2}\left(k-\omega\cos\theta\right)\cos\left[2k\left(t-z\right)\right]\nonumber \\
 & \approx & -{\cal A}^{2}\frac{h^{2}}{z^{2}}\frac{2\omega\left(\omega-k\right)}{\left(\omega-2k\right)}\cos\left[2k\left(t-z\right)\right].\label{eq:-3}
\end{eqnarray}
The amplitude of this term will only be of the same order as $\Delta j_{z,1}(u,v)$
or greater if $\omega\approx2kc=2\omega_{g}$ in SI units, which leads
to the relation:
\begin{eqnarray}
\frac{\left|\Delta j_{z,2}(u,v)\right|_{max}}{j_{z,0}(u,v)}\approx\frac{\omega_{g}h^{2}}{\left(\omega-2\omega_{g}\right)z^{2}} & \gtrapprox & \frac{4h}{z}\approx\left.\frac{\Delta j_{z,1}(u,v)}{j_{z,0}(u,v)}\right|_{\omega=\omega_{g}}.\label{eq:-4}
\end{eqnarray}
For gravitational wave strains at earth of the order of $\nicefrac{h}{z}\sim10^{-21}$
this leads to the condition:
\begin{equation}
\omega-2\omega_{g}\lessapprox\frac{h}{4z}\omega_{g}\sim10^{-22}\omega_{g}\label{eq:time dependent freq. constraint}
\end{equation}
This extreme fine tuning condition for the frequency of the scalar
wave means direct observation of this effect would be impractical.
Therefore, even if we were able to observe a background field with
$\omega\sim2\omega_{g}$ or a gravitational wave source with $\omega_{g}\sim\frac{1}{2}\omega_{CMB}$,
the resonance phenomenon does not lead to an observable effect in
the time dependent oscillation of $j_{z}$.

\subsection{Additive Enhancement of $j_{z}$ \label{subsec:Additive-Enhancement-of}}

The time-averaged current of the scalar plane wave in the $z-$direction
in the absence of a gravitational wave is:
\begin{equation}
\left\langle j_{z}\right\rangle _{h=0}=2\omega\cos\theta{\cal A}^{2}.\label{eq:-5}
\end{equation}
The time-averaged current from (\ref{eq:general ave current}) in
the presence of the gravitational wave is:
\begin{eqnarray}
\left\langle j_{z}\right\rangle  & = & 2\omega\cos\theta{\cal A}^{2}+2{\cal A}^{2}\left(\omega\cos\theta-2k\right)\left(\frac{B_{1}^{2}h^{4}}{z^{4}}+\frac{B_{2}^{2}h^{4}k_{v}^{2}\left(B_{2}k_{v}^{2}z^{2}+2B_{1}\left(B_{5}+B_{4}k_{v}+k_{v}^{2}z^{2}\right)\right)}{z^{2}\left(B_{3}^{2}k_{v}^{2}z^{2}+\left(B_{5}+B_{4}k_{v}+k_{v}^{2}z^{2}\right)^{2}\right)}\right)\nonumber \\
 &  & +{\cal A}^{2}\,\frac{2E^{2}h^{2}k_{v}^{6}z^{4}\left(\omega\cos\theta-k\right)}{k_{v}^{2}z^{2}\left(E_{3}+k_{v}^{2}z^{2}\right)^{2}+\left(E_{4}+E_{2}k_{v}^{2}z^{2}\right)^{2}}\nonumber \\
 &  & -{\cal A}^{2}\,\frac{2E^{2}h^{2}k_{v}^{7}z^{4}\left(3E_{4}k_{v}^{2}z^{2}+E_{2}k_{v}^{4}z^{4}+E_{3}\left(E_{4}-E_{2}k_{v}^{2}z^{2}\right)\right)}{\left(k_{v}^{2}z^{2}\left(E_{3}+k_{v}^{2}z^{2}\right)^{2}+\left(E_{4}+E_{2}k_{v}^{2}z^{2}\right)^{2}\right)^{2}}\,.\label{eq:ave current implications}
\end{eqnarray}
We conclude from this expression that the time-averaged $z-$component
of the scalar four-current is enhanced by the presence of the gravitational
wave. The $z-$component of the scalar four-current gives the number
of scalar particles per unit area flowing through a surface of constant
$z$ per unit time. We interpret the enhancement of $\left\langle j_{z}\right\rangle $
as scalar particle production and not just an apparent increase in
this component of the four-current for the following reason. If this
were only an apparent increase in $j_{z}$, due to expansion and contraction
of space in the $x$ and $y$ directions from the gravitational wave,
it would come as a result of a changing proper three-volume. The proper
3D volume element in Cartesian coordinates for this wave would be:
\begin{equation}
V_{proper}=\sqrt{det\left(g_{ij}\right)}\Delta x\,\Delta y\,\Delta z=f(u,v)\,g(u,v)\,\Delta x\,\Delta y\,\Delta z\approx\left(1-\frac{4h^{2}e^{2iku}}{\left(v-u\right)^{2}}\right)\Delta x\,\Delta y\,\Delta z\label{eq:proper volume}
\end{equation}
which oscillates with time. This means that the time oscillation of
$j_{z}$ discussed in Section \ref{subsec:Sinusoidal-Variation-of}
could be an apparent change in $j_{z}$, but the time independent
enhancement in Eq. (\ref{eq:ave current implications}) cannot be
due to this effect. The assumption that gravitational waves lead to
particle production is not an original conclusion as other authors
have previously made the same assertion \cite{Jones - Scalar Production,Brandenberger}.
The energy that goes into this particle production must come from
the gravitational wave as we can see from Eq'ns (\ref{eq:jx}) and
(\ref{eq:jy}) that the components of the scalar four-currents in
the $x$ and $y$ directions are not affected by the presence of the
gravitational wave.

The enhancements to $\left\langle j_{z}\right\rangle $ are suppressed
by powers of the gravitational strain, $\nicefrac{h}{z}$, which is
very small, but there are two reasons to consider them: 1) they might
lead to observable cumulative effects over very large distances and
2) the resonance effect for $\omega\approx2\omega_{g}$ in the collinear
limit might conceivably make this term relevant. As discussed in Section
\ref{sec:Four-Current}, the enhancement of current is consistent
with the results in \cite{Jones - Scalar Production} concerning $\left\langle j_{u}\right\rangle $
for a plane gravitational wave. The resonance effect, however, is
a new feature due to the distinct nature of the collinear limit in
the spacetime background of a gravitational wave with amplitude that
decays with distance.

Let us first consider the dominant term in (\ref{eq:ave current implications})
which is proportional to $\left(\nicefrac{h}{z}\right)^{2}$. This
effect only exists for scalar waves which are not collinear with the
gravitational wave. As we discussed in Section \ref{sec:Four-Current},
for realistic gravitational wave sources the full form of Eq. (\ref{eq:ave current implications})
is only relevant for extremely small $\theta$. Therefore, we can
take the series expansion of Eq. (\ref{eq:ave current implications})
up to the lowest order term:
\begin{equation}
\left\langle j_{z}\right\rangle _{approx}\approx2\omega\cos\theta{\cal A}^{2}+8{\cal A}^{2}\frac{h^{2}\omega^{2}}{z^{2}k^{2}}\cos^{4}\left(\nicefrac{\theta}{2}\right)\left(\omega\cos\theta-k\right)\cos^{2}\left(2\phi\right)\label{eq:-11}
\end{equation}
where we also used the dominant term in $B_{1}$ from Eq. (\ref{eq:B1soln}).
This effect is largest for $\phi=0,\pm\frac{\pi}{2},\pi$ which corresponds
to scalar waves with off-axial components in the $\pm x$ and $\pm y$
directions.

If we are considering the interaction of the gravitational wave with
a background field analogous to the CMB, this background field should
be isotropic at each point in space. Let us make the simplifying assumption
that an isotropic background field can be treated as a combination
of independent plane waves in every direction, and integrate over
all possible directions to obtain the following expression for the
time-averaged current in the $z-$direction:
\begin{equation}
\left\langle j_{z}\right\rangle _{approx,4\pi}=\int\left\langle j_{z}\right\rangle _{approx}\,d\Omega={\cal A}^{2}\frac{8\pi h^{2}\omega^{2}}{3z^{2}k^{2}}\left(\omega-2k\right).\label{eq:-6}
\end{equation}
The time-averaged $z-$component of the background field itself vanishes
when integrated over all directions as it should. Note that $\left\langle j_{z}\right\rangle _{approx}$
is not actually valid over a small region of solid angle around $\theta=0$,
but this region is so small (see (\ref{eq:collinear angle limits})
and (\ref{eq: numerical angle limit})) that this will not change
the result appreciably. This expression for $\left\langle j_{z}\right\rangle _{approx,4\pi}$
represents an increase in the scalar four-current in the $z-$direction
due to the interaction of the gravitational wave with background scalar
fields which are not propagating in the $z-$direction. Therefore,
this enhancement would be an additive effect since it would not affect
the background field that is involved in producing it.

In order to determine a numerical estimate of the cumulative effect
of this interaction, we need to estimate the size of the constant,
$h$, of the gravitational wave. In order to do this we used the parameters
of the GW150914 event \cite{Abbot et al}. The measured gravitational
wave strain of $\nicefrac{h}{z}\sim10^{-21}$ at an approximate distance
of $z\sim400MPc\approx10^{25}\,m$ leads to an approximate value of
$h\sim10^{4}\,m.$ This value does not seem particularly small, but
it means that the amplitude of the gravitational wave was much less
than 1 for distances $z\apprge1000\,km$ from the source of the gravitational
wave.

For particle production caused by an external field, the particle
production rate per unit volume, $\Gamma$, can be be related to the
current by the relation \cite{Jones - Scalar Production,Frob et al}:
\begin{equation}
\Gamma_{a}\,\Delta T\approx\left\langle j_{z}\right\rangle _{approx,\Omega}\label{eq:-12}
\end{equation}
where $\Delta T$ is a characteristic time scale of the problem. The
characteristic time scale here is $\Delta T\sim\nicefrac{2\pi}{\omega_{g}}.$
This means that the rate of production of particles collinear with
the gravitational wave per unit volume from this effect is:
\begin{eqnarray}
\Gamma_{a} & \approx & \left\langle j_{z}\right\rangle _{approx,\,\Omega}={\cal A}^{2}\frac{\omega_{g}}{2\pi}\frac{8\pi h^{2}\omega^{2}}{3z^{2}\omega_{g}^{2}}\left(\omega-2\omega_{g}\right)={\cal A}^{2}\frac{4h^{2}\omega^{2}}{3z^{2}\omega_{g}}\left(\omega-2\omega_{g}\right).\label{eq:-7}
\end{eqnarray}
where we have converted to SI units. For $z\gg1$ light second, the
rate of production does not change appreciably over a unit distance
so: 
\begin{eqnarray}
\frac{d}{dt}n_{z}=\Gamma_{a} & \Rightarrow & \frac{d}{dz}\left(n_{z}\right)=\frac{\Gamma_{a}}{c}\label{eq:additive enhancement DE}
\end{eqnarray}
which means that:
\begin{equation}
\frac{\Gamma_{a}}{c}={\cal A}^{2}\frac{4h^{2}\omega^{2}}{3cz^{2}\omega_{g}}\left(\omega-2\omega_{g}\right)\label{eq:-13}
\end{equation}
is the rate of increase per unit distance in the $z-$direction of
the number density of scalar particles moving in the $z-$direction,
$n_{z}$. We can convert this to the rate of increase per unit distance
of the number density per unit distance in the $z-$ direction per
unit solid angle about the source of the gravitational wave via:
\begin{equation}
\frac{\Gamma_{a}}{c}\times z^{2}={\cal A}^{2}\frac{4h^{2}\omega^{2}}{3c\omega_{g}}\left(\omega-2k\right).\label{eq:particle production per steradian}
\end{equation}
This expression for the rate of production per steradian is only valid
in the collinear region for a small solid angle about the point of
interaction of the scalar and gravitational waves:
\begin{eqnarray}
\delta\theta=\sqrt{\frac{c}{\omega z}} & \Rightarrow & \delta\Omega\approx\frac{\pi\left(\delta\theta\,R\right)^{2}}{4\pi\,R^{2}}=\frac{c}{4\omega z}\label{eq:solid angle for nz}
\end{eqnarray}
but this solid angle decreases with distance. This means that any
particle moving in a direction within the collinear region at distance
$z$ was definitely moving in a direction within the collinear region
at earlier times when it was closer to the source of the gravitational
wave. The expression in Eq. (\ref{eq:particle production per steradian})
means that within this small solid angular region, $\delta\Omega$,
the particle production rate per steradian per unit distance is constant
with respect to distance from the source. Therefore the cumulative
increase in the number density of scalar particles moving in the $z-$direction
per steradian over distance $\Delta z$ will be given by:
\begin{equation}
{\cal A}^{2}\frac{4h^{2}\omega^{2}}{3c\omega_{g}}\left(\omega-2\omega_{g}\right)\Delta z.\label{eq:-16}
\end{equation}
We can finally determine the total increase in the number density
of particles moving in the $z-$direction at the final distance from
the source:
\begin{eqnarray}
\Delta n_{z}={\cal A}^{2}\frac{4h^{2}\omega^{2}}{3c\omega_{g}}\left(\omega-2\omega_{g}\right)\Delta z\times\frac{1}{z^{2}} & = & {\cal A}^{2}\frac{4h^{2}\omega^{2}}{3c\omega_{g}}\left(\omega-2\omega_{g}\right)\frac{\left(z-z_{0}\right)}{z^{2}}\nonumber \\
 & \approx & {\cal A}^{2}\frac{4h^{2}\omega^{2}}{3cz\omega_{g}}\left(\omega-2\omega_{g}\right)\label{eq:change in nz}
\end{eqnarray}
for $z\gg z_{0}$. Using the parameters of the GW150914 event $\left(z\sim10^{25}m,\,h\sim10^{4}m,\,f_{g}\sim150s^{-1}\right)$
and the parameters of the CMB ($f\approx160\,GHz$) the cumulative
increase in the number density of scalar particles moving in the $z-$direction
relative to the background is:
\begin{equation}
\frac{\Delta n_{z}}{n_{z,0}}=\frac{\Delta n_{z}}{2\omega{\cal A}^{2}}\sim2\times10^{-5}.\label{eq:relative change in nz}
\end{equation}
This is a fairly small effect but should be in the range of detectability. 

\subsection{Exponential Enhancement of $j_{z}$? \label{subsec:Exponential-Enhancement-of}}

In Section \ref{subsec:Additive-Enhancement-of} we discussed the
lowest order enhancement of the scalar current in the $z-$direction
and its possible observable effects. This involved the interaction
of the gravitational wave with noncollinear scalar background fields
which led to an additive enhancement to the scalar current in the
collinear direction.

In this section we will consider the interaction with scalar waves
propagating collinear to the gravitational waves using the collinear
solution for $\Psi$ in Eq. (\ref{eq:collinear solution}). The time-averaged
current in the $z-$direction for a scalar wave propagating collinear
to the gravitational wave is:
\begin{equation}
\left\langle j_{z}\right\rangle _{collinear}=2\omega\cos\theta{\cal A}^{2}+{\cal A}^{2}\frac{\omega^{2}h^{4}}{2\left(\omega-2k\right)z^{4}},\label{eq:collinear ave current implications}
\end{equation}
where, for simplicity, we have ignored the lower order additive enhancement
from the previous section. 

In the previous section we considered the relative increase in the
collinear scalar current due to the interactions with noncollinear
background fields and the normalization of $\Psi$ cancelled out in
these calculations. In this section it is necessary to consider the
normalization of $\Psi$. In relativistic quantum mechanics one of
the possible normalization conventions \cite{Jones - Scalar Production,HalzenMartin}
is to normalize the particle density in the following way :
\begin{equation}
\rho\equiv\left|j_{t}\right|=1\,\mbox{particle per unit volume}.\label{eq:conventional normalization}
\end{equation}
Considering the general result in Eq. (\ref{eq:time dependent jt})
in the limit of no gravitational wave: 
\begin{equation}
\left|j_{t,\,h=0}\right|=2\omega{\cal A}^{2}=1\,m^{-3}\label{eq:-18}
\end{equation}
which leads to the identification ${\cal A}=\sqrt{\nicefrac{1\,m^{-3}}{2\omega}}.$
In order to consider the cumulative effect of the interaction of the
gravitational wave with a background particle field in the universe
we can't choose an arbitrary density of $1\,m^{-3}$. We instead normalize
to the number density, $n_{z}$, of scalar particles associated with
the collinear plane wave using:
\begin{eqnarray}
{\cal A}=\sqrt{\frac{n_{z}}{2\omega}} & \mbox{\qquad}\Rightarrow\mbox{\qquad} & \left|j_{t,\,h=0}\right|=n_{z}.\label{eq:normalization}
\end{eqnarray}
\newpage{}

Using a similar identification as in the last section:

\begin{equation}
\Gamma_{e}\,\Delta T\approx\left\langle \Delta j_{z}\right\rangle _{collinear}\label{eq:-19}
\end{equation}
we arrive at the following expresssion for the production rate per
unit volume of scalar particles moving in the $z-$direction:
\begin{eqnarray}
\Gamma_{e} & = & \frac{1}{\Delta T}{\cal A}^{2}\frac{\omega^{2}h^{4}}{2\left(\omega-2kc\right)z^{4}}=\frac{\omega_{g}}{2\pi}\frac{n_{z}}{2\omega}\frac{\omega^{2}h^{4}}{2\left(\omega-2kc\right)z^{4}}\nonumber \\
 & = & \frac{\omega\,\omega_{g}h^{4}}{8\pi\left(\omega-2\omega_{g}\right)}\frac{n_{z}}{z^{4}}\equiv\Lambda\frac{n_{z}}{z^{4}}\label{eq:exponential production rate-1}
\end{eqnarray}
where we have converted to SI units. The differential equation for
the number density of scalar particles moving in the $z-$direction
is:
\begin{eqnarray}
\frac{d}{dt}n_{z}=\Gamma_{e} & \Rightarrow & c\frac{d}{dz}n_{z}=\Lambda\frac{n_{z}}{z^{4}}.\label{eq:nz diff. eqn}
\end{eqnarray}
Here the rate of increase in $n_{z}$ is dependent on $n_{z}$ which
is different than in Eq. (\ref{eq:additive enhancement DE}) where
the rate of increase did not depend on $n_{z}$. The form of this
differential equation suggests that an exponential enhancement of
$n_{z}$ is possible. The solution to this equation is given by:
\begin{equation}
n_{z}(z)=n_{z,0}\exp\left[\frac{\Lambda}{3c}\left(\frac{1}{z_{0}^{3}}-\frac{1}{z^{3}}\right)\right]\label{eq:nz solution}
\end{equation}
which is exponentially increasing for $z<\sqrt[3]{\frac{\Lambda}{3c}}$,
but for $z>\sqrt[3]{\frac{\Lambda}{3c}}$ rapidly approaches a limiting
value of:
\begin{equation}
n_{z,\infty}=n_{z,0}\exp\left[\frac{\Lambda}{3c\,z_{0}^{3}}\right].\label{eq:limiting nz}
\end{equation}
This formula is only applicable for initial distances where the linearized
Einstein equation is valid. For the GW150914 event at around the time
of peak amplitude the merging black holes had an orbital separation
of approximately $3.5\times10^{5}\,m$ \cite{Basic Physics}. If we
push this analysis as far as it can go, using $z_{0}\sim10^{6}\,m$,
the amplitude of the gravitational wave at this distance would be:
\begin{equation}
\frac{h}{z_{0}}\sim\frac{10^{4}m}{10^{6}m}=10^{-2}\label{eq:strain at z0}
\end{equation}
and the linearized theory would still be valid. Using parameter values
for the GW150914 event and $\omega$ for the peak of the CMB we calculated
the value:
\begin{equation}
\sqrt[3]{\nicefrac{\Lambda}{3c}}\approx750\,m.\label{eq:-20}
\end{equation}
In this situation we have:
\begin{eqnarray}
\sqrt[3]{\nicefrac{\Lambda}{3c}}<h & \& & \sqrt[3]{\nicefrac{\Lambda}{3c}}\ll z_{0}\label{eq:-21}
\end{eqnarray}
which means that there is no exponential growth in the linearized
region, where this analysis is valid, and $n_{z}(z)$ is already very
close to the limiting value at the initial point we have used. The
total increase in the number density of scalar particles relative
to an initial number of particles at $z_{0}\approx10^{6}\,m$ is:
\begin{equation}
\Delta n_{z}=n_{z,\infty}-n_{z,0}\sim10^{-10}\,n_{z0}.\label{eq:-14}
\end{equation}
This shows that the increase in the number density of particles moving
collinear to the gravitational wave from the region where the linearized
theory is valid will not lead to any noticeable effect.

The resonance effect for collinear particles could contribute substantially
to this process since:
\begin{equation}
\sqrt[3]{\nicefrac{\Lambda}{3c}}=\sqrt[3]{\frac{\omega\,\omega_{g}h}{24c\pi\left(\omega-2\omega_{g}\right)}}.\label{eq:-15}
\end{equation}
For $\omega\approx2\omega_{g}$, in order for the exponential region
to extend significantly into the linearized region:
\begin{equation}
\sqrt{\nicefrac{\Lambda}{2c}}>100\,\times h,\label{eq:-24}
\end{equation}
the constraint on the frequency of the scalar wave would be:
\begin{equation}
\left(\omega-2\omega_{g}\right)<\frac{\omega_{g}^{2}\,h}{12\pi c(100)^{2}}\sim\frac{2\pi\left(150s^{-1}\right)\left(10^{4}m\right)}{12\pi\left(3\times10^{8}\nicefrac{m}{s}\right)\left(10^{6}\right)}\omega_{g}\sim1\times10^{-9}\,s^{-1}\omega_{g}.\label{eq:-25}
\end{equation}
This frequency constraint is much less stringent than in Eq. (\ref{eq:time dependent freq. constraint}),
therefore, in the unlikely event that we detect a background field
with $\omega\approx2\omega_{g}$ or a gravitational wave source with
$\omega_{g}\approx\frac{1}{2}\omega_{CMB}$, this resonance effect
might be significant.

\section{Conclusion \label{sec:Conclusion}}

The main result of this paper is the power series solution, given
in Eq'ns (\ref{eq:general solution}) - (\ref{eq:F2 solution}), of
the massless Klein-Gordon equation for a plane wave of a massless
scalar field moving in an arbitrary direction in the spacetime background
of a  gravitational wave with amplitude that decays with distance.
This result is interesting because it is consistent with previous
results \cite{Jones - Scalar Production,Dasgupta - Scalar Fermion}
derived for a plane gravitational wave but has a new and unexpected
behaviour in the limit of collinear scalar and gravitational waves.

In order to clearly illustrate the source of this unexpected collinear
behaviour we presented a detailed discussion of the series solutions
in the collinear limit, Eq. (\ref{eq:collinear solution}), and in
a special noncollinear case, Eq. (\ref{eq:special noncollinear solution})
with coefficients given by (\ref{eq:Asoln}), (\ref{eq:B2soln}),
(\ref{eq:B3soln}), (\ref{eq:B1soln}), (\ref{eq:B5soln}), (\ref{eq:B4soln}),
(\ref{eq:Csoln}) and (\ref{eq:Dsoln}). The special noncollinear
solution was shown to be consistent with the results of \cite{Jones - Scalar Production}
away from the collinear limit. 

One significant new characteristic of the solution in the collinear
limit is the presence of a resonance effect for scalar waves with
frequency close to twice the frequency of the gravitational waves,
$\omega\approx2\omega_{g}$. As we will discuss below, this resonant
behaviour does not lead to any detectable physical effects in the
current analysis with currently known physical parameters. The presence
of this new phenomenon, however, suggests the possiblity of unanticipated
phenomenon in similar analyses for more realistic models. 

In order to discuss the physical implications of these solutions we
have also presented the expressions for the components of the four-current,
$j_{\mu}$, determined from these solutions. The $j_{t}$ and $j_{z}$
components are affected by the presence of the gravitational wave,
but we concentrated on $j_{z}$. There are both time-independent contributions
to $j_{z}$ and contributions which are sinusoidally oscillating with
time. To isolate the time-independent terms of $j_{z}$ we calculated
the time-averaged current in the $z-$direction, $\left\langle j_{z}\right\rangle $.
In the same section we also discussed the range of angles around the
collinear direction for which the exact solution for $\left\langle j_{z}\right\rangle $
is required and the range over which an approximate solution for $\left\langle j_{z}\right\rangle $
is appropriate. This discussion also applies to the range of angles
over which the exact solution in the general case, Eq. (\ref{eq:general solution}),
is necessary. At the distances of currently observed gravitational
wave sources, the range of $\theta$ over which the exact solution
is necessary is very small, but it is necessary to give the correct
collinear limit. The consistency with the results of \cite{Jones - Scalar Production}
is also demonstrated by agreement of our expression for $\left\langle j_{u}\right\rangle $
with theirs away from the collinear region.

In Section \ref{sec:No-Particle-Production} we discussed the collinear
and anti-collinear solutions and used them to obtain a linear combination
from which we can consider the possiblity of a gravitational wave
generating a scalar waveform in a vacuum. The result of this analysis
is that the solution to the KG equation will vanish identically if
you try to eliminate the scalar wave that exists independently of
the gravitational wave. From this we conclude that the gravitational
wave can only interact with, and enhance the four-current of, a scalar
plane wave and there is no scalar particle production in a vacuum.
This conclusion is consistent with the results of \cite{Dasgupta - Scalar Fermion,Gibbons}.

We discusssed three different types of physical implications of the
solutions in this gravitational wave model. The first effect was time-dependent
sinusoidal variations in the scalar current in the $z-$direction,
$j_{z}$. The largest effect of this type was only present in the
noncollinear case and predicted a variation with relative amplitude
of the order of $10^{-6}$. This result is consistent with the predictions
of \cite{Dasgupta - Scalar Fermion} and while somewhat small, should
be in the range of detectability. In addition, we showed that the
resonance effect in the collinear limit is extremely unlikely to lead
to a detectable signal.

The second type of physical effect we considered was the time-independent
enhancement of the time-averaged current $\left\langle j_{z}\right\rangle $
relative to the current in the absence in the gravitational wave.
We argued that this enhancement could not be an apparent increase
in the current because the effect of the gravitational wave on the
three-volume would be sinsuoidally varying and would be eliminated
by the time averaging. Therefore we conclude, as did the authors of
\cite{Jones - Scalar Production}, that this enhancement correponds
to scalar particle production as a result of the interaction with
the gravitational wave with the scalar plane wave. We calculated that
the interaction of the gravitational wave with the noncollinear components
of an isotropic background scalar field with a frequency of the order
of $\omega\approx\omega_{CMB}$ could lead to an enhancement of the
density of scalar particles moving in the $z-$direction of the order:
\begin{equation}
\frac{\Delta n_{z}}{n_{0,z}}\sim2\times10^{-5}.\label{eq:-26}
\end{equation}
This is a fairly small effect but it might be detectable.

The final physical effect we discussed was the possibility that the
interaction of the gravitational wave with the collinear scalar current
could lead to an exponential enhancement of the collinear current.
Unfortunately, for the parameter range of observed gravitational waves,
the exponential enhancement only occurs very close to the source where
the linearized theory will not be applicable. In these cases, the
interaction of the gravitational wave with the collinear component
of a background scalar field does not lead to a detectable effect.
The resonance effect in the collinear case would only help if we could
observe a scalar wave with $\omega\approx2\omega_{g}$ or a gravitational
wave source with $\omega_{g}\approx\frac{1}{2}\omega_{CMB}$ which
are far outside the range of theoretical predictions and/or practical
detectability.

As we discussed in the introduction, the results of this paper were
derived for a background scalar field which, as far as we know, does
not exist in our current universe. This was done in order to examine
specific details of the interaction of a quantum field with a gravitational
wave with spatially dependent amplitude in a paper of reasonable length.
There are a number of avenues for future research. We would like to
extend this analysis to fermionic and EM fields to look for new features
and/or different physical predictions. Another area for future research
is to include the time dependence of a transient gravitational wave
pulse since this could be much more significant for the strongest
gravitational wave sources. We would also like to include a more realistic
model for the background radiation field than the \emph{ad hoc }model
of independent scalar plane waves in every direction. Finally, it
would be very interesting to extend the gravitational wave background
beyond the solutions to the linearized Einstein equation.

\subsection*{Appendix A}

The terms up to order $1/z^{4}$ resulting from substituting the assumed
form in Eq. (\ref{eq:general solution}) in the Klein-Gordon equation
(\ref{eq:our DE}) are:

\begin{eqnarray}
\mbox{indep. of }h\,e^{iku}: &  & \left(-2A\,k_{v}+\frac{k_{x}^{2}}{2}+\frac{k_{y}^{2}}{2}\right)d_{1}\left(u,v\right)^{2}d_{2}\left(u,v\right)^{2}d_{3}\left(u,v\right)\nonumber \\
 &  & \equiv\alpha\,d_{1}\left(u,v\right)^{2}d_{2}\left(u,v\right)^{2}d_{3}\left(u,v\right),\label{eq:-27}\\
\nonumber \\
\frac{1}{(v-u)}\:\mbox{terms}: &  & -\frac{2he^{iku}}{(v-u)}\,\left(2E\,k\,k_{v}+k_{x}^{2}-k_{y}^{2}-E\,\alpha\right)\,\equiv\,-\frac{2he^{iku}}{(v-u)}\:\beta,\label{eq:-28}\\
\nonumber \\
\frac{e^{iku}}{(v-u)^{2}}\:\mbox{terms}: &  & -\frac{4ihe^{iku}}{(v-u)^{2}}\left[\left(E\left(A+k-k_{v}\right)+2E_{2}k\right)-E_{2}\frac{\alpha}{k_{v}}+\left(2\left(B_{3}-E_{2}\right)-E_{2}F\right)\frac{\beta}{k_{v}}\right]\nonumber \\
 &  & \equiv-\frac{4ihe^{iku}}{(v-u)^{2}}\:\gamma,\label{eq:-29}\\
\nonumber \\
\frac{e^{2iku}}{(v-u)^{2}}\:\mbox{terms}: &  & \frac{4h^{2}e^{2iku}\,}{(v-u)^{2}}k_{v}\left[2\left(3A+k\right)-4\left(B_{1}+B_{2}\right)k+2E^{2}k+\left(1+B_{1}+B_{2}-E^{2}\right)\frac{\alpha}{k_{v}}-E\,\frac{\beta}{k_{v}}\right]\nonumber \\
 &  & \equiv\,\frac{4h^{2}e^{2iku}}{(v-u)^{2}}\:\delta,\label{eq:-30}\\
\nonumber \\
\frac{e^{iku}}{(v-u)^{3}}\:\mbox{terms}: &  & \frac{8he^{iku}}{(v-u)^{3}}\left[\frac{E}{k_{v}}\left(2E_{3}k+E_{2}\left(A+k-k_{v}\right)-k_{v}\right)-E\,E_{3}\frac{\alpha}{k_{v}^{2}}+\left(2B_{3}-E_{2}-F_{2}\right)\frac{\gamma}{k_{v}}\right.\nonumber \\
 &  & \left.\hspace{0.6in}-\left(\left(B_{3}-E_{2}\right)^{2}+2\left(B_{3}^{2}+E_{3}+B_{4}k_{v}+B_{5}\right)-\left(2B_{3}-E_{2}-F_{2}\right)F_{2}\right)\frac{\beta}{k_{v}^{2}}\right]\nonumber \\
 &  & \equiv\frac{8he^{iku}}{(v-u)^{3}}\:\zeta,\label{eq:-31}
\end{eqnarray}
\begin{eqnarray}
\frac{e^{2iku}}{(v-u)^{3}}\:\mbox{terms}: &  & -\frac{8ih^{2}e^{2iku}}{(v-u)^{3}}\left[2\left(B_{1}+B_{2}\right)\left(A+2k-k_{v}\right)-A+k_{v}-2k\left(2B_{2}B_{3}-2Ck_{v}+E^{2}E_{2}\right)\right.\nonumber \\
 &  & \left.\hspace{0.9in}+E\,E_{2}\frac{\beta}{k_{v}}+\left(2E_{2}+F_{2}-2B_{3}\right)\frac{\delta}{k_{v}}+\left(B_{2}B_{3}-Ck_{v}+E^{2}E_{2}\right)\frac{\alpha}{k_{v}}\right]\\
 &  & \equiv-\frac{8ih^{2}e^{2iku}}{(v-u)^{3}}\,\epsilon,\\
\\
\frac{e^{3iku}}{(v-u)^{3}}\:\mbox{terms}: &  & \frac{8h^{3}e^{3iku}}{(v-u)^{3}},k_{v}\left[\left\{ 2\left(B_{1}+B_{2}\right)E\,k-6Fk+6E\left(A+k\right)\right\} k_{v}-\left(E\left(B_{1}+B_{2}-1\right)-F\right)\frac{\alpha}{k_{v}}\right.\\
 &  & \left.\hspace{0.9in}-\left(B_{1}+B_{2}\right)\frac{\beta}{k_{v}}\right]\nonumber \\
 &  & \equiv\frac{8h^{3}e^{3iku}}{(v-u)^{3}}\:\theta,\label{eq:-33}\\
\nonumber \\
\frac{e^{iku}}{(v-u)^{4}}\:\mbox{terms}: &  & \frac{16ihe^{iku}}{(v-u)^{4}}\left[\frac{E}{k_{v}^{2}}\left\{ E_{3}\left(A+k-k_{v}\right)\frac{\;}{\;}\hspace{-0.3cm}-2\left(E_{2}E_{3}+E_{4}\right)k-E\,E_{2}k_{v}\right\} +\left(2B_{3}-F_{2}\right)\frac{\zeta}{k_{v}}\right.\nonumber \\
 &  & \hspace{0.9in}+E\left(2E_{2}E_{3}+E_{4}\right)\frac{\alpha}{k_{v}^{3}}-\left(2\left(B_{3}^{2}+B_{4}k_{v}+B_{5}\right)+\left(B_{3}-F_{2}\right)^{2}\right)\frac{\gamma}{k_{v}^{2}}\nonumber \\
 &  & \hspace{1.2in}+\left\{ 2\left(E_{2}E_{3}+E_{4}\right)+2B_{3}\left(2B_{3}^{2}+3\left(B_{4}k_{v}+B_{5}\right)\right)\begin{array}{c}
\,\\
\,
\end{array}\right.\nonumber \\
 &  & \hspace{1.2in}\left.\left.\begin{array}{c}
\,\\
\,
\end{array}-F_{2}\left(2\left(B_{3}^{2}+B_{4}k_{v}+B_{5}\right)+\left(B_{3}-F_{2}\right)^{2}\right)\right\} \frac{\beta}{k_{v}^{3}}\right],\label{eq:-34}\\
\nonumber \\
\frac{e^{2iku}}{(v-u)^{4}}\:\mbox{terms}: &  & -\frac{16h^{2}e^{2iku}}{(v-u)^{4}}\left[3\left(A+2k-k_{v}\right)\left(\frac{B_{2}B_{3}}{k_{v}}-C\right)+3\left(B_{1}+B_{2}\right)\right.\nonumber \\
 &  & \hspace{0.5in}+2k\left(E^{2}\left(E_{2}^{2}+E_{3}\right)-2B_{2}\left(B_{3}^{2}+B_{4}k_{v}+B_{5}\right)\right)-E\left(E_{2}^{2}+E_{3}\right)\frac{\beta}{k_{v}^{2}}\nonumber \\
 &  & \hspace{0.5in}+\left(B_{2}\left(B_{3}^{2}+B_{4}k_{v}+B_{5}\right)-E^{2}\left(E_{2}^{2}+E_{3}\right)\right)\frac{\alpha}{k_{v}^{2}}+\left(2E_{2}+F_{2}-2B_{3}\right)\frac{\epsilon}{k_{v}}\label{eq:-35}\\
 &  & \hspace{0.5in}-\left(2\left(B_{3}E_{2}+E_{3}+B_{4}k_{v}+B_{5}\right)+3\left(B_{3}-E_{2}\right)^{2}-\left(2B_{3}-2E_{2}-F_{2}\right)F_{2}\right)\frac{\delta}{k_{v}^{2}},\nonumber \\
\nonumber \\
\frac{e^{3iku}}{(v-u)^{4}}\:\mbox{terms}: &  & -\frac{16ih^{3}e^{3iku}}{(v-u)^{4}}\left[2E\,k\left(B_{2}B_{3}-Ck_{v}+\left(B_{1}+B_{2}\right)E_{2}\right)-3\left(E-F\right)\left(A+3k-k_{v}\right)\right.\nonumber \\
 &  & \hspace{1.1in}+5E\,k-6F\left(E_{2}-F_{2}\right)k-\left(B_{2}B_{3}-Ck_{v}+\left(B_{1}+B_{2}\right)E_{2}\right)\frac{\beta}{k_{v}}\nonumber \\
 &  & \hspace{1.1in}+\left(F\left(E_{2}-F_{2}\right)-E\left(B_{2}B_{3}-Ck_{v}+\left(B_{1}+B_{2}\right)E_{2}\right)\right)\frac{\alpha}{k_{v}}\nonumber \\
 &  & \hspace{1.1in}\left.\left(E_{2}+F_{2}-2B_{3}\right)\frac{\theta}{k_{v}}\right],\label{eq:-36}\\
\frac{e^{4iku}}{(v-u)^{4}}\:\mbox{terms}: &  & \frac{16h^{4}e^{4iku}}{(v-u)^{4}}k_{v}\left[2\left(B_{1}+B_{2}\right)\left(3A+5k\right)-2\left(A+k+\left(4D-E\,F\right)k\right)-F\frac{\beta}{k_{v}}\right.\nonumber \\
 &  & \hspace{1in}\left.+\left(B_{1}+B_{2}+D-E\,F\right)\frac{\alpha}{k_{v}}\right],\label{eq:-37}
\end{eqnarray}
where:
\begin{eqnarray}
d_{1}\left(u,v\right) & = & \left(1+\frac{2iB_{3}}{k_{v}\left(v-u\right)}+\frac{4\left(B_{4}k_{v}+B_{5}\right)}{k_{v}^{2}\left(v-u\right)^{2}}\right),\label{eq:-38}\\
d_{2}\left(u,v\right) & = & \left(1-\frac{2iE_{2}}{k_{v}\left(v-u\right)}+\frac{4E_{3}}{k_{v}^{2}\left(v-u\right)^{2}}-\frac{8iE_{4}}{k_{v}^{3}\left(v-u\right)^{3}}\right),\label{eq:-39}\\
d_{3}\left(u,v\right) & = & \left(1-\frac{2iF_{2}}{k_{v}\left(v-u\right)}\right).\label{eq:-40}
\end{eqnarray}

\subsection*{Appendix B}

The full expression, up to order $\nicefrac{1}{z^{4}}$, for the scalar
four-current in the $z-$direction is:
\begin{eqnarray}
j_{z}(u,z) & = & 2{\cal A}^{2}\omega\cos\theta+2{\cal A}^{2}\left({\cal B}^{2}+{\cal C}^{2}\right)\left(\omega\cos\theta-2k\right)+2{\cal A}^{2}\left({\cal D}^{2}+{\cal E}^{2}\right)\left(\omega\cos\theta-k\right)\nonumber \\
 &  & +2{\cal A}^{2}\left({\cal D}\,\partial_{z}{\cal E}-{\cal E}\,\partial_{z}{\cal D}\right)+2\left[{\cal D}\left(k-2\omega\cos\theta\right)-\partial_{z}{\cal E}\right]\sin\left(ku\right)\nonumber \\
 &  & +2{\cal A}^{2}\left[{\cal E}\left(k-2\omega\cos\theta\right)+\partial_{z}{\cal D}\right]\cos\left(ku\right)+2\left[2{\cal B}\left(k-\omega\cos\theta\right)+\partial_{z}{\cal C}\right]\sin\left(2ku\right)\nonumber \\
 &  & -2{\cal A}^{2}\left[2{\cal C}\left(k-\omega\cos\theta\right)-\partial_{z}{\cal B}\right]\cos\left(2ku\right)+2\left[{\cal G}\left(3k-2\omega\cos\theta\right)-\partial_{z}{\cal F}\right]\sin\left(3ku\right)\nonumber \\
 &  & +2\left[{\cal F}\left(3k-2\omega\cos\theta\right)+\partial_{z}{\cal G}\right]\cos\left(3ku\right)+4{\cal A}^{2}{\cal H}\left(\omega\cos\theta-2k\right)\cos\left(4ku\right)\label{eq:time dependent jz}
\end{eqnarray}
where:
\begin{eqnarray}
{\cal H} & = & \frac{h^{4}}{8k^{2}z^{4}}\left(9A^{2}+16Ak+3k^{2}\right)+\frac{Eh^{4}}{8kz^{4}}\left(3EA+5Ek+2Fk\right),\nonumber \\
{\cal B} & = & \frac{Ch^{2}}{z^{3}}-\frac{B_{2}B_{3}h^{2}k_{v}^{3}z}{B_{3}^{2}k_{v}^{2}z^{2}+\left(B_{4}k_{v}+B_{5}+k_{v}^{2}z^{2}\right)^{2}},\nonumber \\
{\cal C} & = & \frac{B_{1}h^{2}}{z^{2}}+\frac{B_{2}h^{2}k_{v}^{2}\left(B_{4}k_{v}+B_{5}+k_{v}^{2}z^{2}\right)}{B_{3}^{2}k_{v}^{2}z^{2}+\left(B_{4}k_{v}+B_{5}+k_{v}^{2}z^{2}\right)^{2}},\nonumber \\
{\cal D} & = & \frac{E\,hk_{v}^{3}z^{2}\left(E_{4}+E_{2}k_{v}^{2}z^{2}\right)}{k_{v}^{2}z^{2}\left(E_{3}+k_{v}^{2}z^{2}\right)^{2}+\left(E_{4}+E_{2}k_{v}^{2}z^{2}\right)^{2}},\label{eq:-41}\\
{\cal E} & = & \frac{E\,hk_{v}^{4}z^{3}\left(E_{3}+k_{v}^{2}z^{2}\right)}{k_{v}^{2}z^{2}\left(E_{3}+k_{v}^{2}z^{2}\right)^{2}+\left(E_{4}+E_{2}k_{v}^{2}z^{2}\right)^{2}},\nonumber \\
{\cal F} & = & -\frac{Fh^{3}k_{v}^{2}}{z\left(\left(F_{2}\right)^{2}+k_{v}^{2}z^{2}\right)},\nonumber \\
{\cal G} & = & \frac{F\,F_{2}\,h^{3}k_{v}}{z^{2}\left(\left(F_{2}\right)^{2}+k_{v}^{2}z^{2}\right)},\nonumber 
\end{eqnarray}
and: 
\begin{equation}
k_{v}=-\omega\frac{\left(1-\cos\theta\right)}{2}.\label{eq:-42}
\end{equation}

The full expression, up to order $\nicefrac{1}{z^{4}}$, for the scalar
four-current in the $t-$direction is:
\begin{eqnarray}
j_{t}(u,z) & = & -2{\cal A}^{2}\omega+2{\cal A}^{2}\omega\left({\cal B}^{2}+{\cal C}^{2}\right)\left(2k-\omega\right)+2{\cal A}^{2}\left({\cal D}^{2}+{\cal E}^{2}\right)\left(k-\omega\right)\nonumber \\
 &  & -2{\cal D}\left(k-2\omega\right)\sin\left(ku\right)-2{\cal E}\left(k-2\omega\right)\cos\left(ku\right)\nonumber \\
 &  & -4{\cal B}\left(k-\omega\right)\sin\left(2ku\right)+4{\cal A}^{2}{\cal C}\left(k-\omega\right)\cos\left(2ku\right)\nonumber \\
 &  & -2\left[{\cal G}\left(3k-2\omega\right)\right]\sin\left(3ku\right)-2\left[{\cal F}\left(3k-2\omega\right)\right]\cos\left(3ku\right)\nonumber \\
 &  & +4{\cal A}^{2}{\cal H}\left(2k-\omega\right)\cos\left(4ku\right).\label{eq:time dependent jt}
\end{eqnarray}

\section*{Acknowledgements}

The authors would like to acknowledge the Brandon University Research
Committee, Brandon University/Brandon University Students Union Workstudy
Program and the Canada Summer Jobs Program for providing funding enabling
JE to work during the summer as a Research Assistant.

\end{document}